\documentclass[ twoside,journal]{IEEEtran}
    \usepackage{makecell}
    \usepackage{array}
    \usepackage{graphicx,amssymb,amsmath}
    \usepackage{multicol}
    \usepackage[noadjust]{cite}
    \usepackage{setspace}
    \usepackage{subfigure}
    \usepackage{graphicx}
    \usepackage{float}
    \usepackage {url}
    \usepackage{stfloats}
    \usepackage{amsthm,pifont}
    \usepackage{flushend}
    \usepackage{cases,subeqnarray}
    \usepackage{bm,multirow,bigstrut}
    \usepackage{amsmath, amsthm, amssymb}
    \usepackage{textcomp}
    \usepackage{latexsym,bm}
    \usepackage{booktabs}
    \usepackage{xcolor}
    \usepackage{mathtools}
    \usepackage{dsfont}
    \usepackage{extarrows}
    \usepackage{epsfig}
    \usepackage{epsfig}
    \usepackage{epstopdf}
    \usepackage[noend]{algpseudocode}
    \usepackage{hyperref}
    \usepackage{algorithmicx,algorithm}

    \theoremstyle{plain}

    \theoremstyle{plain}

    \usepackage{amsmath}

\begin{document}
    \title{Generative AI based Secure Wireless Sensing for ISAC Networks}

  \author{Jiacheng Wang, Hongyang Du, Yinqiu Liu, Geng Sun, Dusit~Niyato,~\IEEEmembership{Fellow,~IEEE}, Shiwen Mao,~\IEEEmembership{Fellow,~IEEE}, Dong In Kim,~\IEEEmembership{Fellow,~IEEE}, and Xuemin~(Sherman)~Shen,~\IEEEmembership{Fellow,~IEEE}

       \thanks{Jiacheng~Wang, Yinqiu Liu, and Dusit Niyato are with the School of Computer Science and Engineering, Nanyang Technological University, Singapore 639798 (e-mail: jiacheng.wang@ntu.edu.sg, yinqiu001@ntu.edu.sg, dniyato@ntu.edu.sg).}
       \thanks{Hongyang Du is with the Department of Electrical and Electronic Engineering, University of Hong Kong, Pok Fu Lam, Hong Kong (e-mail: duhy@eee.hku.hk).}
       \thanks{Geng Sun is with College of Computer Science and Technology, Jilin University, China 130012, (e-mail: sungeng@jlu.edu.cn).}
       \thanks{S. Mao is with the Department of Electrical and Computer Engineering, Auburn University, Auburn, USA (e-mail: smao@ieee.org).}
        \thanks{Dong In Kim is with the Department of Electrical and Computer Engineering, Sungkyunkwan University, Suwon 16419, South Korea (email:dongin@skku.edu).}
        \thanks{X. Shen is with the Department of Electrical and Computer Engineering, University of Waterloo, Canada (e-mail: sshen@uwaterloo.ca).}
    


      }
    
    \maketitle
    
    \begin{abstract}
        Integrated sensing and communications (ISAC) is expected to be a key technology for 6G, and channel state information (CSI) based sensing is a key component of ISAC. However, current research on ISAC focuses mainly on improving sensing performance, overlooking security issues, particularly the unauthorized sensing of users. In this paper, we propose a secure sensing system (DFSS) based on two distinct diffusion models. Specifically, we first propose a discrete conditional diffusion model to generate graphs with nodes and edges, guiding the ISAC system to appropriately activate wireless links and nodes, which ensures the sensing performance while minimizing the operation cost. Using the activated links and nodes, DFSS then employs the continuous conditional diffusion model to generate safeguarding signals, which are next modulated onto the pilot at the transmitter to mask fluctuations caused by user activities. As such, only ISAC devices authorized with the safeguarding signals can extract the true CSI for sensing, while unauthorized devices are unable to achieve the same sensing. Experiment results demonstrate that DFSS can reduce the activity recognition accuracy of the unauthorized devices by approximately 70\%, effectively shield the user from the unauthorized surveillance.      
    \end{abstract}
    \IEEEpeerreviewmaketitle
    \section{Introduction}
    Integrated Sensing and Communication (ISAC) is an emerging technology, which integrates wireless sensing and communication into a single system, effectively utilizing the network resources to simultaneously perform data exchange and environmental sensing~\cite{cui2021integrating}. Such integration not only enhances the spectral and energy efficiency, but also reduces operational and hardware costs, offering broad applications in scenarios such as the internet of vehicles~\cite{sun2023bargain, zheng2021uav}. A key example of ISAC is the CSI-based sensing~\cite{kong2024csi}, which has seen rapid advancement over the last decade. This technology involves the analysis of channel state information (CSI) from wireless communications networks to sense humans, including their locations, activities, and even breathing and heart rates~\cite{liu2019wireless}.

\begin{figure*}[t]
	\centering
\includegraphics[width=0.95\textwidth]{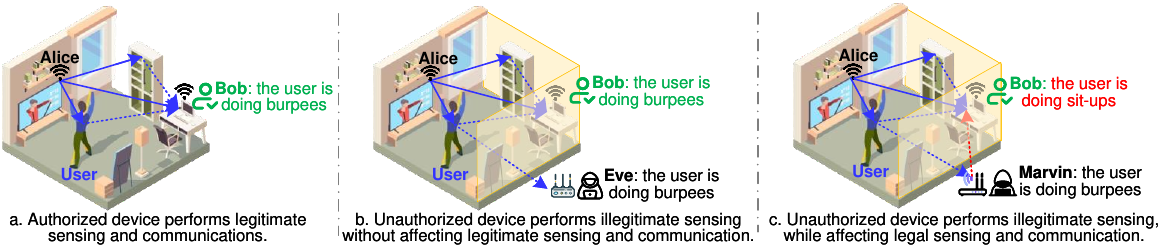}%
	\caption{The principles of CSI based sensing and two main types of security issues. Here, Alice is the ISAC device which transmits the wireless signal and Bob is the authorized device, which performs legitimate sensing of the user. Eve is an unauthorized eavesdropper who performs illegitimate sensing of the user by capturing and processing the signals. Marvin is an unauthorized attacker who can not only sense the users, but also generate spoofing signals to disrupt legitimate communications and sensing.}
	\label{CASE}
\end{figure*}

    In CSI-based sensing, the research is comprehensive but reveals major security weaknesses. Rigorously, these systems rely on the CSI, which is derived based on the communication protocols and training symbols shared between the signal transmitter and receiver in wireless communications networks~\cite{wang2024unified}. This implies that any standard wireless device in open spaces can intercept these signals and use the standardized training symbols to measure the CSI. By analyzing CSI measurements, unauthorized devices can then use these CSI measurements to glean insights into an individual's daily activities~\cite{avola2022person}. For example, an unauthorized access point (AP) could capture Wi-Fi signals and extract CSI to deduce an individual's gait, posing a substantial privacy risk~\cite{wang2024acceleration}. Another major concern is the vulnerability to attacks from rogue APs, such as using customized APs to capture and alter signal amplitude and phase to create spoofing signals. These spoofing signals are transmitted to disrupt legitimate sensing activities, adversely impacting the performance of the ISAC system~\cite{liu2023time}. Figure~\ref{CASE} depicts two primary security challenges in CSI sensing systems, emphasizing the crucial need to protect individuals' daily activities from unauthorized AP monitoring.

    To shield users from unauthorized surveillance, an effective approach is to generate a safeguarding signal and modulate it onto the training symbols for CSI estimation, thereby masking the signal fluctuations caused by the user activity. Here, it is crucial that this safeguarding signal remains unreconstructable by unauthorized devices. Moreover, the signal generation must adapt to changes in external conditions, which presents a significant challenge. Fortunately, the recent progress in artificial intelligence generated content (AIGC)~\cite{lin2024blockchain} has led to the development of generative AI (GAI) models, providing support for the signal generation~\cite{wang2024generative}.

    Within the field of GAI, diffusion models have gained particular attention due to their versatility and effectiveness. Diffusion models can generate text, images, and videos based on user prompts~\cite{lin2023unified}, providing solid support for applications like DALL-E. Furthermore, diffusion models excel in signal denoising~\cite{wang2024generativeJ} and generation~\cite{chi2024rf}, delivering many unique advantages. First, they demonstrate creativity, capable of producing data that resembles yet differs from the training data, thereby ensuring variety. Second, they generate data through a denoising, which relies on specific network parameters and random seeds, leading to unpredictable results. Furthermore, these models exhibit considerable adaptability, implying that they can fine-tune the generation process to match user prompts closely~\cite{du2024enhancing}. These characteristics make diffusion models a suitable choice for safeguarding signal generation.

    Building on the discussion above, we propose a diffusion model based secure sensing system (DFSS). Specifically, we first propose a discrete conditional diffusion model to generate the graph with nodes and edges. This graph guides the ISAC system to activate wireless links appropriately for sensing, ensuring performance while minimizing link activation for cost-effective operation. Then, the DFSS employs a continuous conditional model to generate safeguarding signals. Compared to methods based on predefined algorithms and codebook design, the generated safeguarding signals show greater randomness and diversity, making them harder to replicate. These signals are then modulated onto the pilot symbols at the transmitter to conceal fluctuations caused by user activities. At the receiver, authorized devices use the same model and seed to generate identical safeguarding signals, enabling them to extract real CSI for communication and sensing. However, unauthorized devices are unable to generate the safeguarding signal, and hence fail to extract CSI for sensing. This effectively protects users from illegitimate monitoring. In summary, the contributions of this paper are as follows.

    \begin{itemize}
    \item We propose a discrete conditional diffusion model (D-CDM) to generate graphs by using ISAC device distribution and user locations as conditions. These graphs guide appropriate activation of wireless links, ensuring optimal sensing performance while minimizing the costs for efficient ISAC network operation.

    \item We design safeguarding signals tailored to human indoor activity patterns and develop a dataset to train a continuous diffusion model. This enables the generation of safeguarding signals based on activated links and the user to be protected. The generated signals exhibit diversity and randomness, making them challenging for unauthorized devices to replicate.

    \item We propose to modulate the generated safeguarding signals onto the pilot at the transmitter. This prevents unauthorized receivers from extracting the actual CSI, which carries detailed information about the surrounding environment and users, thereby shielding users from illegitimate sensing.

    \item We evaluate the DFSS using software-defined radio devices. Taking activity recognition as an example, the results demonstrate that DFSS can reduce the activity recognition accuracy of the state-of-the-art system by about 70\%, confirming its effectiveness in protecting users from illegitimate sensing.
    
\end{itemize}
    
    \section{Related Work}
    In this section, we review existing CSI based sensing systems and analyze the security issues within them.
    \subsection{CSI Based Sensing}
    With the proliferation of wireless devices, CSI-based sensing technologies have been significantly advanced. Researchers have proposed various CSI-based sensing systems, including localization~\cite{wang2023through,wang2018deep}, activity and gesture recognition~\cite{wang2024unified,ahmed2020device}, respiratory and heartbeat monitoring~\cite{wang2021smartphone,wang2017tensorbeat}, and authentication~\cite{lin2023contactless}. In~\cite{wang2016csi}, the authors trained the weights of the deep network as fingerprints and employed a greedy learning to reduce complexity. Then, they used a radial basis function-based probabilistic approach to achieve localization, with a mean error of approximately 0.94 m. The authors in~\cite{wang2024generative} employed clustering to analyze the multi-domain parameters of reflections caused by humans. This facilitated the human flow detection, especially the number of subflows and subflow sizes, achieving the detection accuracy of 92\% and 91\%, respectively. For activity recognition, the authors in~\cite{chen2018wifi} proposed an attention-based bi-directional long short-term memory network to learn features from sequential CSI measurements in two directions, achieving an overall recognition accuracy of 97.3\%. For small scale sensing, the authors in~\cite{wang2017tensorbeat} derived breathing signals from the CSI phase difference between antennas, and then integrated the signal matching algorithm with the peak detection to estimate respiratory rates. In single-user scenarios, evaluations show an estimation error of less than 0.5 bpm in 96\% of cases.
    \subsection{Security Issues in CSI Based Sensing}
    In a CSI sensing system, there are two main types of security issues. \textbf{The first type} involves unauthorized devices performing illegitimate sensing to obtain user's relevant information~\cite{li2016csi}, as shown in part $b$ of the Fig.~\ref{CASE}. For instance, the authors in~\cite{lu2024imperceptible} proposed a signal propagation model that converts eavesdropped CSI into those of legitimate devices. Then, they added user identification models into the substitute model set for training the signal pattern calibration generative model, enabling the eavesdropping of user activity with an accuracy of up to 80\%. Given that various hand coverage and finger movements create unique interference to signals, the authors in~\cite{meng2019revealing} extracted temporal and frequency domain fluctuations of CSI and employed dynamic time warping to achieve keystroke recognition, allowing eavesdrop on user passwords. Evaluations show that the system achieves an average classification accuracy of 93.5\%.
 \begin{figure*}[t]
	\centering
\includegraphics[width=1\textwidth]{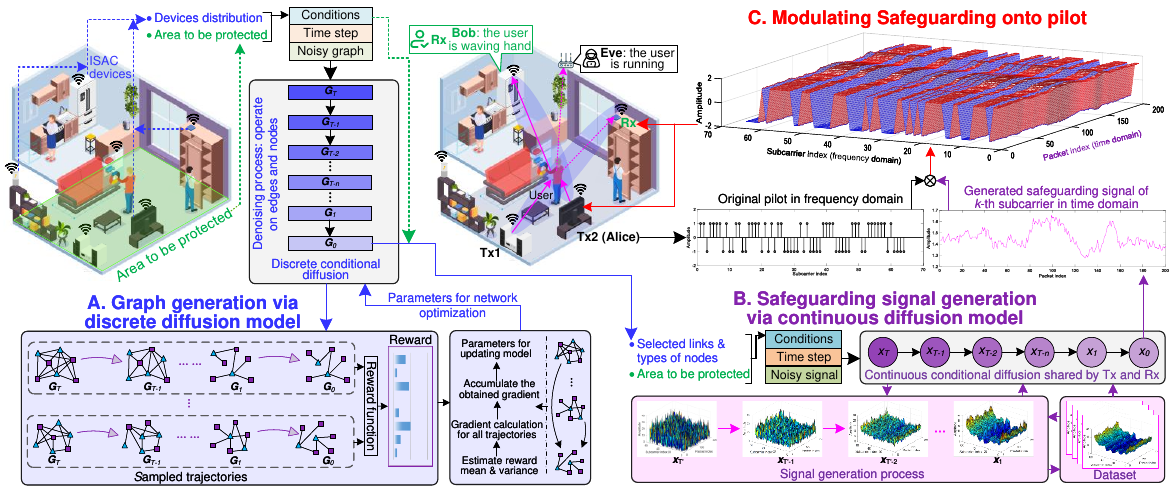}%
	\caption{The framework of the proposed DFSS. DFSS consists of three primary modules, including graph generation, safeguarding signal generation, and signal modulation. The first module generates the graph with nodes and edges based on the locations of users and the distribution of ISAC devices, which directs the activation of sensing links. Subsequently, the second module produces safeguarding signals by utilizing the generated graphs and user locations. Finally, the third module modulates these safeguarding signals onto a pilot to mask the CSI amplitude fluctuations triggered by user activities. Through this process, authorized devices can use the safeguarding signals to extract actual CSI from the captured signals for legitimate wireless sensing, whereas unauthorized devices are unable to do so, thereby preventing illegitimate sensing.}
	\label{FRWK}
\end{figure*}

    \textbf{The second type} is to attack the legitimate sensing, leading the authorized system to make wrong decisions~\cite{liu2023time}, as shown in part $c$ of the Fig.~\ref{CASE}. For example, the authors in~\cite{zhou2022wiadv} introduced a signal synthesis scheme to generate adversarial signals with specific motion features. Then it employed a black-box attack strategy to address discrepancies between the perturbation space and the classifier’s input space, misleading the sensing system into incorrect gesture classification. Experimental results demonstrate that the average attack success rate of this system is over 70\%. The authors in~\cite{xu2022wicam} developed an adversarial perturbation to degrade sensing performance, while also designing a mask to confine the perturbation to targeted areas, minimizing its impact on communication. Evaluations indicate that it can reduce the accuracy of the sensing model to below 50\%, while decreasing the impact of perturbation on bit error rate by 77.78\%.

    The previous discussion underscores the extensive research focus on sensing users illicitly, gathering information about their location and behavior without their knowledge. To safeguard users’ daily activities from unauthorized monitoring, this paper introduces DFSS. It generates and modulates a safeguarding signal onto the pilot signal to mask the signal fluctuations caused by the user activity, thereby preventing unauthorized devices from illegitimate sensing.
   
    \section{System Design}
    \subsection{System Overview}
    The system consists of three main modules, as shown in Fig.~\ref{FRWK}. First, by taking the spatial distribution of ISAC devices and user location as inputs, we propose D-CDM to generate a graph, which guides the activation of wireless links, as well as signal transmitters (Txs) and receivers (Rxs) for sensing the user. Next, we use a continuous conditional diffusion model to generate a safeguarding signal given the activated links, and locations of Txs, Rxs, and user. Finally, the Tx modulates the generated safeguarding signal onto the pilot and then transmits it for user sensing. In this way, the signal fluctuations caused by user activities are masked by the modulated safeguarding signal. Therefore, only authorized Rx can use the same diffusion model, inputs, and random seed to regenerate the safeguarding signal, thereby extracting true CSI data for effective sensing and communication. Unauthorized devices, however, are unable to reproduce the same signal for CSI extraction and user sensing, therefore shielding users from unauthorized surveillance. Figure 2 provides the overall system structure.

    \subsection{Signal Model}
    Consider an ISAC link that includes an orthogonal frequency division multiplexing (OFDM) signal transmitter and a receiver. When the receiver capture signals, it estimates the CSI using the pilot shared between the transmitter and receiver for sensing and communication. Assuming there is no inter-carrier interference (ICI)~\cite{biguesh2006training}, the training symbols for the $N$ subcarriers can be expressed as a diagonal matrix ${\bf{X}} = {\rm{diag}}\left( {X\left[ 0 \right], \  \ldots, \ X\left[ {N - 1} \right]} \right)$, where $X\left[ n \right]$ indicates the pilot signal on the $n$-th subcarrier. Let $H\left[ n \right]$ be channel gain of the $n$-th subcarrier, then the received training symbol can be expressed as ${\bf{Y}} \buildrel \Delta \over = {\bf{X}} \times {\bf{H}}$, where ${\bf{H}} = \left[ {H\left[ 0 \right], \ldots ,H\left[ {N - 1} \right]} \right]$. Based on ${\bf{Y}}$ and the predefined training symbol ${\bf{X}}$, ${\bf{H}}$ can be estimated by minimizing the following cost function
\begin{equation}
\begin{aligned}\label{eq1}
J\left( {{\bf{\hat H}}} \right) &= {\left\| {{\bf{Y}} - {\bf{X\hat H}}} \right\|^2} \\ 
&= {\left( {{\bf{Y}} - {\bf{X\hat H}}} \right)^{\rm{H}}}\left( {{\bf{Y}} - {\bf{X\hat H}}} \right)\\ 
 &= {{\bf{Y}}^H}{\bf{Y}} - {{\bf{Y}}^H}{\bf{X\hat H}} - {{{\bf{\hat H}}}^H}{{\bf{X}}^H}{\bf{Y}} + {{{\bf{\hat H}}}^H}{{\bf{X}}^H}{\bf{X\hat H}}.
\end{aligned}
\end{equation}
Let 
\begin{align}\label{eq2}
{{\partial J\left( {{\bf{\hat H}}} \right)} \mathord{\left/
 {\vphantom {{\partial J\left( {{\bf{\hat H}}} \right)} {\partial {\bf{\hat H}}}}} \right.
 \kern-\nulldelimiterspace} {\partial {\bf{\hat H}}}} =  - 2{\left( {{{\bf{X}}^H}{\bf{Y}}} \right)^ * } + 2{\left( {{{\bf{X}}^H}{\bf{X\hat H}}} \right)^ * } = 0,
\end{align}
then the channel estimation can be obtained
\begin{align}\label{eq3}
{\bf{\hat H}} = {\left( {{{\bf{X}}^H}{\bf{X}}} \right)^{ - 1}}{{\bf{X}}^H}{\bf{Y}} = {{\bf{X}}^{ - 1}}{\bf{Y}}.
\end{align}
For the $n$-th subcarrier, the CSI at time $t$ can be expressed as
\begin{align}\label{eq4}
H\left( {{f_n},t} \right) &= {e^{ - j\varepsilon }}{H_s}\left( {{f_n},t} \right)\\ \notag
 &+ {e^{ - j\varepsilon }}\sum\limits_{l \in {P_d}} {\alpha_l \left( {{f_n},t} \right){e^{{{ - j2\pi {f_n}{d_l}\left( t \right)} \mathord{\left/
 {\vphantom {{ - j2\pi {f_n}{d_l}\left( t \right)} c}} \right.
 \kern-\nulldelimiterspace} c}}}}  + {n_{{f_n},t}}, 
\end{align}
where ${e^{ - j\varepsilon }}$ is the phase offset, ${H_s}\left( {{f_n},t} \right)$ is the sum of CSIs for static propagation paths, ${P_d}$ is the set of dynamic paths caused by the moving user, ${\alpha _l}\left( {{f_n},t} \right)$ is a complex number including the attenuation and initial phase of the $l$-th
path, ${e^{{{ - j2\pi {f_n}{d_l}\left( t \right)} \mathord{\left/
 {\vphantom {{ - j2\pi {f_n}{d_l}\left( t \right)} t}} \right.
 \kern-\nulldelimiterspace} c}}}$ is the phase shift caused by the change of the $l$-th dynamic path, and ${n_{{f_n},t}}$ is the noise. 
 
 Based on (\ref{eq4}), we can obtain ${\left| {H\left( {{f_n},t} \right)} \right|^2}$, shown as~(\ref{eq5}) at the bottom of this page. 
 \newcounter{mycount}
    \begin{figure*}[b]
	\normalsize
	\setcounter{mycount}{\value{equation}}
	\hrulefill
	\vspace*{4pt}
    \begin{align}\label{eq5}
\begin{array}{c}
{\left| {H\left( {{f_n},t} \right)} \right|^2} = \sum\limits_{l \in {P_d}} {{{\left| {\alpha_l \left( {{f_n},t} \right)} \right|}^2}}  + {\left| {{H_s}\left( {{f_n},t} \right)} \right|^2} + \sum\limits_{l \in {P_d}} {2\left| {{H_s}\left( {{f_n},t} \right){\alpha _l}\left( {{f_n},t} \right)} \right|\cos \left[ {{{2\pi {f_n}\left( {{v_l}t + {d_l}\left( 0 \right)} \right)} \mathord{\left/
 {\vphantom {{2\pi {f_n}\left( {{v_l}t + {d_l}\left( 0 \right)} \right)} c}} \right.
 \kern-\nulldelimiterspace} c} + {\theta _{sl}}} \right]} \\ 
 \qquad + \sum\limits_{l,l' \in {P_d}} {\left( {2\left| {{\alpha _l}\left( {{f_n},t} \right){\alpha _{l'}}\left( {{f_n},t} \right)} \right|\left. { \times \cos \left( {{{2\pi {f_n}\left( {\Delta {v_{ll'}}t + \Delta {d_{ll'}}\left( 0 \right)} \right)} \mathord{\left/
 {\vphantom {{2\pi {f_n}\left( {\Delta {v_{ll'}}t + \Delta {d_{ll'}}\left( 0 \right)} \right)} c}} \right.
 \kern-\nulldelimiterspace} c} + {\theta _{ll'}}} \right)} \right) + \Psi \left( {{n_{{f_n},t}}} \right)} \right.} .
\end{array}
    \end{align} 
    \end{figure*}
 Here, $\Psi \left( {{n_{{f_n},t}}} \right)$ represents the power of the cross terms that are multiplied by noise, ${v_l}$ is the rate of length change of the $l$-th dynamic path, ${d_l}\left( 0 \right)$ is the initial length, ${\theta _{sl}}$ and ${\theta _{ll'}}$ denote initial phases, $\Delta {v_{ll'}} = {v_l} - {v_{l'}}$ is the rate difference between two paths, and $\Delta {d_{ll'}}\left( 0 \right) = {d_l}\left( 0 \right) - {d_{l'}}\left( 0 \right)$ represents the initial length difference. From (\ref{eq5}), we can observe that the CSI power is composed of a series of constants and sinusoids, with the overall intensity dependent on $\left| {{H_s}\left( {{f_n},t} \right)} \right|$. Additionally, the overall fluctuation characteristics of the CSI are primarily determined by the third term with the intensity smaller than ${\left| {{H_s}\left( {{f_n},t} \right)} \right|^2}$, which is a constant. As different user behaviors result in various ${v_l}$, the user behavior recognition can be achieved by extracting and analyzing fluctuation characteristics. To mitigate fluctuations caused by user behaviors, this paper proposes to generate the safeguarding signal $s\left( {{f_n},t} \right)$, which is then multiplied with the training symbols used for CSI estimation at the transmitter. Hence, based on ${\bf{X}}$, we have 
    \begin{align}\label{eq6}
    {\bf{X'}} = {\rm{diag}}\left( {s\left( {{f_0},t} \right)X\left[ 0 \right], \ldots ,s\left( {{f_{N - 1}},t} \right)X\left[ {N - 1} \right]} \right).
\end{align}
The generated protection signal is shared only between authorized transmitters and receivers. Therefore, based on~(\ref{eq1}), for the authorized receiver, we have ${\bf{\hat H}} = {\left( {{\bf{X'}}} \right)^{ - 1}}{\bf{Y'}}$, where ${\bf{Y'}}$ is the signal received when ${\bf{X'}}$ is transmitted. For the unauthorized receiver, ${\bf{\hat H'}} = {{\bf{X}}^{ - 1}}{\bf{Y'}}$. On this basis, according to (\ref{eq4}) and (\ref{eq5}), we have 
\begin{align}\label{eq7}
{\left| {H'\left( {{f_n},t} \right)} \right|^2} = {\left| {s\left( {{f_n},t} \right)} \right|^2}{\left| {H\left( {{f_n},t} \right)} \right|^2},
\end{align}
where 
\begin{align}\label{eq8}
H'\left( {{f_n},t} \right) &= {e^{ - j\varepsilon }}s\left( {{f_n},t} \right){H_s}\left( {{f_n},t} \right)  + {{n'}_{{f_n},t}}\\ \notag
 &+ {e^{ - j\varepsilon }}\sum\limits_{l \in {P_d}} {s\left( {{f_n},t} \right)\alpha_l \left( {{f_n},t} \right){e^{{{ - j2\pi {f_n}{d_l}\left( t \right)} \mathord{\left/
 {\vphantom {{ - j2\pi {f_n}{d_l}\left( t \right)} c}} \right.
 \kern-\nulldelimiterspace} c}}}}.
\end{align}

For example, if the CSI power is ${\left| {{H_{wh}}\left( {{f_n},t} \right)} \right|^2}$ when the user waves hand, then the safeguarding signal can be set as ${1 \mathord{\left/
 {\vphantom {1 {\left| {{H_{wh}}\left( {{f_n},t} \right)} \right|}}} \right.
 \kern-\nulldelimiterspace} {\left| {{H_{wh}}\left( {{f_n},t} \right)} \right|}}$, thereby the CSI power received by unauthorized receivers is $1 = {1 \mathord{\left/
 {\vphantom {1 {{{\left| {{H_{wh}}\left( {{f_n},t} \right)} \right|}^2}}}} \right.
 \kern-\nulldelimiterspace} {{{\left| {{H_{wh}}\left( {{f_n},t} \right)} \right|}^2}}} \times {\left| {{H_{wh}}\left( {{f_n},t} \right)} \right|^2}$. In this way, the safeguarding signal effectively masks signal fluctuation features caused by user gestures. Here, the fluctuation characteristics are determined by several parameters, mainly include ${P_d}$, ${v_l}$, and ${d_l}\left( 0 \right)$. While the overall intensity is primarily determined by the static path components in ${H_{wh}}\left( {{f_n},t} \right)$, which are directly related to the locations of the signal transmitter and receiver. However, in practice, it is challenging to obtain the aforementioned parameters, making it impossible to calculate ${1 \mathord{\left/
 {\vphantom {1 {\left| {{H_{wh}}\left( {{f_n},t} \right)} \right|}}} \right.
 \kern-\nulldelimiterspace} {\left| {{H_{wh}}\left( {{f_n},t} \right)} \right|}}$ directly. 
 
Given the strong generation capabilities of GAI, hence, this paper first proposes D-CDM, which generates graph based on the given ISAC device distribution and user location to guide the activation of wireless links and signal Txs and Rxs. On this basis, the continuous diffusion model is utilized to generate the safeguarding signal by using the activated links, nodes, and user location as conditions. This signal is then modulated onto training symbols to mask the signal fluctuations caused by user activities, thereby providing user with protection against unauthorized surveillance.  
 
\subsection{Graph Generation for Link Selection}
Given the distribution of ISAC devices and the user location, we first determine the sensing link and signal transmitter and receiver. Treating signal transmitters and receivers as nodes, sensing links as edges, and the user location and device distribution as conditions, this paper proposes D-CDM to generate a graph to guide the ISAC system to activate appropriate links and devices, ensuring effective and economical system operation.
\subsubsection{Problem Formulation}
Based on (\ref{eq6}) to (\ref{eq8}), we can observe that the dynamic path contains all the information related to user activity. Therefore, the sensing signal-to-noise ratio (SSNR), which is the ratio of the dynamic path power to the sum of the noise power and the static path power, can be used to characterize the sensing performance
\begin{align}\label{eq9}
SSNR = \frac{{\sum\limits_{l \in {P_d}} {{{\left| {\alpha_l \left( {{f_n},t} \right)} \right|}^2}} }}{{{{\left| {{H_s}\left( {{f_n},t} \right) + {n_{{f_n},t}}} \right|}^2}}}.
\end{align}
Here, a higher SSNR implies better sensing performance. According to~\cite{wang2022placement}, given the signal transceiver pair, the SSNR can be further expressed as
\begin{align}\label{eq10}
SSNR = \frac{{\vartheta \sigma }}{{4\pi {{\left( {{d_{Tx}}{d_{Rx}}} \right)}^2}\left( {{{\gamma \vartheta } \mathord{\left/
 {\vphantom {{\gamma \vartheta } {d_D^2}}} \right.
 \kern-\nulldelimiterspace} {d_D^2}} + b} \right)}},
\end{align}
where $\vartheta $ is positively correlated with both signal transmission power and antenna gain, $\sigma $ represents the effective reflection area of the user, ${d_{Tx}}$ and ${d_{Rx}}$ denote the distances between the user and the signal transmitter and receiver, respectively, $\gamma$ and $b$ are two constants for a given transceiver pair and can be measured in advance, and ${d_D}$ is the path length of line-of-sight (LoS) path. As can be seen from (\ref{eq10}), the sensing performance depends on the distance from the user to the signal transmitter and receiver. Given the spatial distribution of devices and user location, it is essential for an ISAC service provider to ensure the sensing performance while activating fewer wireless links. Thereby, the problem of selecting appropriate sensing links, Txs, and Rxs can be formulated as the following optimization problem
\begin{equation}
\begin{aligned}\label{eq11}
\mathop {\max }\limits_{\bf{G}} \{ &{U_{LK}} = {\alpha_1}SSNR_T - {\alpha_2}\# \left( {Tx} \right) - {\alpha_3}\# \left( {Link} \right) \} \\  
s.t.,\  &\# \left( {{\rm{node}}} \right) > \# \left( {Tx} \right) \ge 1, \\ 
& \# \left( {{\rm{node}}} \right) > \# \left( {Rx} \right) \ge 1,\\
{\rm{ }} & \# \left( {Tx} \right) \times \left\{ {\# \left( {{\rm{node}}} \right) - \# \left( {Tx} \right)} \right\} \ge \# \left( {Link} \right) \ge 1,\\ 
{\rm{        }} &Tx \ne Rx,\\
 &Tx \nleftrightarrow T'x,\\
  &Rx \nleftrightarrow R'x,\\
\end{aligned}
\end{equation}
where $SSNR_T$ is the sum of the SSNR for all links, ${\bf{G}}$ is the matrix that describes the link and node activation, ${\alpha _1}$, ${\alpha _2}$, and ${\alpha _3}$ are weighting factors, $\# \left( {Tx} \right)$, $\# \left( {Rx} \right)$, $\# \left( {Link} \right)$, and $\# \left( {node} \right)$ denote the number of signal transmitters, receiver, links, and the total number of ISAC devices, respectively, $Tx \nleftrightarrow Tx'$ means no link among transmitters, and $Rx \nleftrightarrow Rx'$ indicates no link between receivers.
\subsubsection{Graph generation}
To solve the above optimization problem, we propose D-CDM to generate the optimal ${{\bf{G}}_0}$ for activation. The D-CDM includes a forward diffusion process and a reverse diffusion process. Unlike diffusion models for images, which add and remove noise at each pixel, D-CDM operates on the nodes and edges. Therefore, the state space consists of node types and edge types. Let ${{{k}}_i}$ represent the one-hot encoding~\cite{liu2024graph} of the attributes for node $i$, then by organizing encodings into a matrix, we obtain ${\bf{K}}$. Similarly, for edges, ${\bf{E}}$ is used to group the one-hot encoding ${{{e}}_{ij}}$ of each edge. On this basis, the optimal edge and node selection scheme can be expressed as ${{\bf{G}}_0} = \left( {{{\bf{{\bf{K}}}}_0},{{\bf{E}}_0}} \right)$. For any node and edge, the transition probabilities at the $t'$-th step are defined as
\begin{align}\label{eq12}
{\left[ {{\bf{Q}}_{t'}^K} \right]_{ij}} = q\left( {{k^{t'}} = j\left| {{k^{t' - 1}} = i} \right.} \right)
\end{align}
and 
\begin{align}\label{eq13}
{\left[ {{\bf{Q}}_{t'}^E} \right]_{i'j'}} = q\left( {{e^{t'}} = j'\left| {{e^{t' - 1}} = i'} \right.} \right),
\end{align}
respectively, where $i,j \in \left\{ {1, \ldots ,a} \right\}$ and $i',j' \in \left\{ {1, \ldots ,a'} \right\}$ are the types of nodes and edges, respectively. Based on these transition matrices, the forward diffusion process is 
\begin{equation}
\begin{aligned}\label{eq14}
q\left( {{{\bf{G}}_T}\left| {{{\bf{G}}_0}} \right.} \right) &= \prod\limits_{t' = 1}^T {q\left( {{{\bf{G}}_T}\left| {{{\bf{G}}_{t' - 1}}} \right.} \right)}  \\
&= \prod\limits_{t' = 1}^T {\left( {{{\bf{{\bf{K} }}}_0}{\bf{Q}}_{t'}^K,{{\bf{E}}_0}{\bf{Q}}_{t'}^E} \right)} \\ 
 &= \left( {{{\bf{{\bf{K} }}}_0}{\bf{\bar Q}}_{t'}^K,{{\bf{E}}_0}{\bf{\bar Q}}_{t'}^K} \right),
\end{aligned}    
\end{equation}
where $T$ is the total number of noise addition steps. 

Besides forward diffusion, another part of D-CDM is the conditional denoising network parameterized by $\theta $. Using the distribution of ISAC devices and user location as conditions and the noisy graph as input, the denoising network can generate a clear graph through denoising, which maximize ${U_{LK}}$ as defined in \ref{eq11}. Rigorously, This denoising process is 
\begin{align}\label{eq15}
{p_\theta }\left( {{{\bf{G}}_0}\left| {{{\bf{G}}_T}} \right.,{\bf{D}}} \right) = p\left( {{{\bf{G}}_T}} \right)\prod\limits_{t' = 1}^T {{p_\theta }\left( {{{\bf{G}}_{t' - 1}}\left| {{{\bf{G}}_{t'}}} \right.,{\bf{D}}} \right)},
\end{align}
where ${\bf{D}}$ is the generation condition, and 
\begin{equation}\label{eq16}
\begin{aligned}
{{p_\theta }\left( {{{\bf{G}}_{t' - 1}}\left| {{{\bf{G}}_{t'}}} \right.,{\bf{D}}} \right)} &= \underbrace {\prod\limits_{1 \le i \le a} {{p_\theta }\left( {k_{t' - 1}^i\left| {{{\bf{G}}_{t'}}} \right.,{\bf{D}}} \right)} }_{node}\\ 
 &\times \underbrace {\prod\limits_{1 \le i',j' \le a} {{p_\theta }\left( {e_{t' - 1}^{i'j'}\left| {{{\bf{G}}_{t'}}} \right.,{\bf{D}}} \right)} }_{edge}.
\end{aligned}
\end{equation}
For each term of a node, we have 
\begin{equation}
\begin{aligned}\label{eq17}
&{p_\theta }\left( {k_{t' - 1}^i\left| {{{\bf{G}}_{t'}}} \right.,{\bf{D}}} \right) \\ 
&= \int_{{k^i}} {{p_\theta }\left( {k_{t' - 1}^i\left| {{k^i}} \right.,{{\bf{G}}_{t'}},{\bf{D}}} \right)} d{p_\theta }\left( {{k^i}\left| {{{\bf{G}}_{t'}}} \right.,{\bf{D}}} \right)\\ 
 &= \sum\limits_{k \in A} {{p_\theta }\left( {k_{t' - 1}^i\left| {{k^i}} \right. = k,{{\bf{G}}_{t'}},{\bf{D}}} \right)} \hat p_K^i\left( k \right),
\end{aligned}
\end{equation}
where $A$ is the space of categories for nodes\footnote{The categories of nodes should be defined according to a specific application. In this paper, nodes are mainly classified into two categories, including signal transmitters and signal receivers.}. Similarly, for each term of an edge, we have 
\begin{equation}
\begin{aligned}\label{eq18}
&{p_\theta }\left( {e_{t' - 1}^{i'j'}\left| {{{\bf{G}}_{t'}}}  \right.,{\bf{D}}} \right)\\  &= \sum\limits_{e \in A'} {{p_\theta }\left( {e_{t' - 1}^{i'j'}\left| {{e^{i'j'}}} \right. = e,{{\bf{G}}_{t'}},{\bf{D}}} \right)} \hat p_E^{i'j'}\left( e \right),
\end{aligned}
\end{equation}
where $A'$ is the space of categories for edges.

Based on the constructed denoising network, we need to further refine its network parameters to facilitate the D-CDM in generating the optimal graph for sensing links and nodes activation. By using ${U_{LK}}$ as the reward function, here the goal is to maximize the expected reward over the sample distribution
\begin{align}\label{eq19}
{J_{ER}}\left( \theta  \right) = {\mathbb{E}_{{{\bf{G}}_0} \sim {p_\theta }\left( {{{\bf{G}}_0}} \right)}}\left[ {r\left( {{{\bf{G}}_0}} \right)} \right],
\end{align}
where $r( \cdot )$ is the reward function defined by (\ref{eq11}). However, optimizing ${J_{ER}}\left( \theta  \right) $ directly is challenging as the likelihood ${p_\theta }\left( {{{\bf{G}}_0}} \right)$ is unavailable. Therefore, we model the conditional denoising process as a Markov decision process. In this process, the agent interacts with its environment through multiple steps. At each step, the agent first observes the current state and then selects an action based on its policy. Following this, the agent receives a reward and transitions to a new state according to the transition function. Applying the above principle to the denoising network, we have
\begin{equation}\label{eq20}
    \begin{aligned}
\left\{ \begin{array}{l}
{{\bf{s}}_{t'}} \buildrel \Delta \over = \left( {{{\bf{G}}_{T - t'}},T - t'} \right)\\
{{\bf{a}}_{t'}} \buildrel \Delta \over = {{\bf{G}}_{T - t' - 1}}\\
{\pi _\theta }\left( {{{\bf{a}}_{t'}}\left| {{{\bf{s}}_{t'}}} \right.} \right) \buildrel \Delta \over = {p_\theta }\left( {{{\bf{G}}_{T - t' - 1}}\left| {{{\bf{G}}_{T - t'}}} \right.} \right)\\
r\left( {{{\bf{s}}_{t'}},{{\bf{a}}_{t'}}} \right) \buildrel \Delta \over = \left\{ {\begin{array}{*{20}{c}}
{r\left( {{{\bf{G}}_0}} \right),{\rm{ }}\ t' = T{\rm{ }}}\\
{0\qquad {\rm{    , }}\ t' < T,}
\end{array}} \right.{\rm{ }}
\end{array} \right.
    \end{aligned}
\end{equation}
where ${{\bf{s}}_{t'}}$ and ${{\bf{a}}_{t'}}$ represent the state and action at the $t'$-th step, respectively, ${\pi _\theta }$ is the policy corresponding to the reverse transition distribution, and $r\left( {{{\bf{s}}_{t'}},{{\bf{a}}_{t'}}} \right)$ is the reward. In this way, the graph generation process, denoted as $\left\{ {{{\bf{G}}_{T}},{{\bf{G}}_{T - 1}}, \ldots ,{{\bf{G}}_0}} \right\}$, can be regarded as a state-action trajectory produced by agent actions within the Markov process ${\bf{\varsigma }} = \left\{ {({{\bf{s}}_0},{{\bf{a}}_0}),({{\bf{s}}_1},{{\bf{a}}_1}), \ldots ,({{\bf{s}}_{T}},{{\bf{a}}_{T}})} \right\}$. According to \ref{eq20}, we have $\sum\nolimits_{t' = 0}^{T} {r\left( {{{\bf{s}}_{t'}},{{\bf{a}}_{t'}}} \right)}  = r\left( {{{\bf{G}}_0}} \right)$. Hence, the expected reward of the agent can be denoted as 
\begin{align}\label{eq21}
{J_{C}}\left( \theta  \right) = {\mathbb{E}_{p\left( {{\bf{\varsigma }}\left| {{\pi _\theta }} \right.} \right)}}\left[ {r\left( {\bf{\varsigma }} \right)} \right] = {\mathbb{E}_{{p_\theta }\left( {{{\bf{G}}_{0:T}}} \right)}}\left[ {r\left( {{{\bf{G}}_0}} \right)} \right],
\end{align}
which is equivalent to ${J_{ER}}\left( \theta  \right)$. On this basis, according to the policy-based learning principle~\cite{liu2024graph}, the gradient boosting is used to optimize the policy gradient. Given ${J_{C}}\left( \theta  \right) =  {\mathbb{E}_{{p_\theta }\left( {{{\bf{G}}_{0:T}}} \right)}}\left[ {r\left( {{{\bf{G}}_0}} \right)} \right]$, we have
\begin{align}\label{eq22}
{\nabla _\theta }{J_{C}}\left( \theta  \right) = {{{\rm \mathbb{E}}}_{\bf{\varsigma }}}\left[ {r\left( {{{\bf{G}}_0}} \right)\sum\limits_{t' = 1}^{T} {{\nabla _\theta }\log {p_\theta }\left( {{{\bf{G}}_{t' - 1}}\left| {{{\bf{G}}_{t'}}} \right.} \right)} } \right],
\end{align}
which is generally intractable. Inspired by DDPO~\cite{black2023training}, which uses reinforcement learning to estimate the policy gradient and yields impressive outcome of image generation, we adopt Monte Carlo estimation to approximate (\ref{eq22}), obtaining
\begin{align}\label{eq23}
{\nabla _\theta }{J_C}\left( \theta  \right) \approx \frac{{T\sum\limits_{z \in Z} {\sum\limits_{t' \in \Gamma } {r\left( {{\bf{G}}_0^{\left( z \right)}} \right){\nabla _\theta }\log {p_\theta }\left( {{\bf{G}}_{t' - 1}^{\left( z \right)}\left| {{\bf{G}}_{t'}^{\left( z \right)}} \right.} \right)} } }}{{\left| Z \right|\left| \Gamma  \right|}},
\end{align}
where $Z$ and $\Gamma $ represent the sets of sampled trajectories and timesteps\footnote{Here, $\left| Z \right| = 256$, $\Gamma $ is an integer randomly sampled from 1 to 30.}, respectively, and ${\bf{G}}_{0:T}^{\left( z \right)},{\rm{ }}z = 1, \ldots ,Z$ are $Z$ trajectories sampled from ${p_\theta }\left( {{{\bf{G}}_{0:T}}} \right)$. Essentially, the policy gradient estimation in (\ref{eq23}) is a weighted summation of gradients, which can produce fluctuating and unreliable policy gradient estimates when the number of Monte Carlo samples is limited. To address this issue, (\ref{eq23}) is further modified into
\begin{align}\label{eq24}
{\nabla _\theta }{J'_C}\left( \theta  \right) = \frac{{T\sum\limits_{z \in Z} {\sum\limits_{t' \in \Gamma } {r\left( {{\bf{G}}_0^{\left( z \right)}} \right){\nabla _\theta }\log {p_\theta }\left( {{\bf{G}}_0^{\left( z \right)}\left| {{\bf{G}}_{t'}^{\left( z \right)}} \right.} \right)} } }}{{\left| Z \right|\left| \Gamma  \right|}}.
\end{align}

As can be observed, the potential number of graph trajectories is vast. Therefore, we can group them into different equivalent classes where trajectories with the same ${{\bf{G}}_0^{\left( z \right)}}$ are considered equivalent. In this way, the number of these classes is less than the total number of graph trajectories, and the optimization can be performed over the classes, which is simpler than optimizing in the entire trajectory space. At last, ${\nabla _\theta }{J'_C}\left( \theta  \right)$ is used to update the network parameters via $\theta ' = \theta  + {\nabla _\theta }{J'_C}\left( \theta  \right) \times \eta $, where $\eta $ represents the learning rate. Upon completing the training, the denoising network can generate graphs based on new input conditions to guide the activation of links and nodes.

\begin{figure}[t]
\centering
\includegraphics[height=4.8cm]{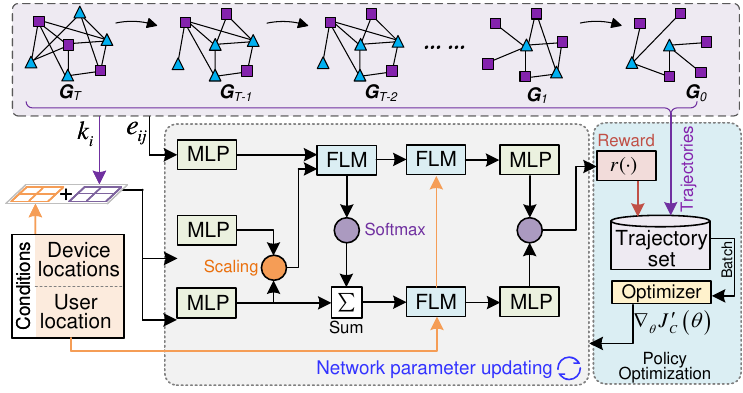} 
\caption{The design of the denoising network and optimization process.} 
\label{NTW} 
\end{figure}

To facilitate learning of graph generation strategies by the denoising network while considering the generation conditions, we build the denoising network based on the graph transformer architecture~\cite{dwivedi2020generalization}, and the overall structure is demonstrated in Fig.~\ref{NTW}. Specifically, the input nodes and edges are first processed by a multi-layer perceptron (MLP) module~\cite{vignac2023digress} to produce embeddings. These embeddings are then passed through a graph transfer layer that features an attention mechanism. Similar to conventional ones, the scoring at each attention layer is defined by feature-wise linear modulation, which can adjust the output of the transformer layer through a scaling and shifting operation based on condition ${\bf{D}}$. Following this, the embeddings of nodes and edges undergo residual connections and layer normalization. These steps prevent the vanishing of gradients and ensure a stable training process. Finally, two separate MLPs are used to decode the node and edge embeddings, respectively, leading to the matrices corresponding to the prediction for ${{\bf{G}}_0}$.

\subsection{Safeguarding Signal Generation}
Using the generated ${{\bf{G}}_z^0}$, combined with user locations as conditions, we further train the continuous diffusion model to produce the safeguarding signal. Unlike the previous method, which optimized the graph generation process using the reward function, here, to guarantee performance, we train the continuous diffusion model with a dataset. This enables the model to generate a safeguarding signal that is similar to yet distinct from the one used in training, thereby ensuring diversity and randomness.

\subsubsection{Continuous Diffusion Model}
Similar to discrete diffusion models, continuous conditional diffusion models also incorporate a forward process and a reverse process. Given a schedule of noise scales $0 < {\beta _1}, \ldots ,{\beta _{T'}} < 1$, for any given safeguarding training sample ${{\bf{x}}_0} \sim {q'_{sg}}\left( {\bf{x}} \right)$, the forward process perturbs the training sample by adding the noise over $T'$ steps, such that 
\begin{align}\label{eq25}
q'\left( {{{\bf{x}}_{t'}}\left| {{{\bf{x}}_{t' - 1}}} \right.} \right) = {{\cal N}}\left( {{{\bf{x}}_{t'}};\sqrt {1 - {\beta _{t'}}} {{\bf{x}}_{t' - 1}},{\beta _{t'}}{\bf{I}}} \right),
\end{align}
where ${\bf{I}}$ is the identity matrix. Based on (\ref{eq25}), we have 
\begin{align}\label{eq26}
{q'_{{\alpha _{t'}}}}\left( {{{\bf{x}}_{t'}}\left| {{{\bf{x}}_0}} \right.} \right) = {\cal N}\left( {{{\bf{x}}_{t'}};\sqrt {{\alpha _{t'}}} {{\bf{x}}_0},\left( {1 - {\alpha _{t'}}} \right){\bf{I}}} \right),
\end{align}
where ${\alpha _{t'}} = \prod\nolimits_{t'' = 1}^{t'} {\left( {1 - {\beta _{t''}}} \right)} $. After the addition of noise, the disrupted data distribution can be denoted as
\begin{align}\label{eq27}
{q'_{{\alpha _{t'}}}}\left( {{\bf{\hat x}}} \right) = \int {{{q'}_{sg}}\left( {\bf{x}} \right)} {q'_{{\alpha _{t'}}}}\left( {{\bf{\hat x}}\left| {\bf{x}} \right.} \right)d{\bf{x}}.
\end{align}
In the forward process described above, the noise ${\beta _1}, \ldots ,{\beta _{T'}}$ is prescribed, allowing ${{\bf{x}}_{T'}}$ to approximate ${{\cal N}}\left( {{\bf{0}},{\bf{I}}} \right)$.

The reverse process is defined as a Markov chain, which is parameterized by 
\begin{equation}
\begin{aligned}\label{eq28}
&{ p'_{\theta '}}\left(  {{{\bf{x}}_{t' - 1}}\left| {{{\bf{x}}_{t'}},{\bf{D'}}} \right.} \right) \\
&= {\cal N}\left( {{{\bf{x}}_{t' - 1}};\frac{{{{\bf{x}}_{t'}} + {\beta _{t'}}{{\bm{\mu }}_{{\bf{\theta '}}}}\left( {{{\bf{x}}_{t'}},t',{\bf{D'}}} \right)}}{{\sqrt {1 - {\beta _{t'}}} }},{\beta _{t'}}{\bf{I}}} \right),
\end{aligned} 
\end{equation}
where ${\bf{D'}}$ is the generation conditions. The objective here is to train the denoising network to generate the safeguarding signal from the noise based on ${\bf{D'}}$, and ensure that the distribution of the generated signal is consistent with that of the training samples. Therefore, the denoising network is trained using a re-weighted version of the evidence lower bound shown in (\ref{eq29}) at the bottom of the next page. 
    \begin{figure*}[b]
	\normalsize
	\setcounter{mycount}{\value{equation}}
	\hrulefill
	\vspace*{4pt}
    \begin{align}\label{eq29}
\theta '' = \mathop {\arg \min }\limits_{\theta '} \sum\limits_{t' = 1}^{T'} {\left( {1 - {\alpha _{t'}}} \right)} {\mathbb{E}_{{{q'}_{data}}\left( {\bf{x}} \right)}}{\mathbb{E}_{{{q'}_{{\alpha _{t'}}}}\left( {{\bf{\hat x}}\left| {\bf{x}} \right.} \right)}}\left[ {\left\| {{{\bm{\mu }}_{{\bf{\theta '}}}}\left( {{\bf{\hat x}},t',{\bf{D'}}} \right) - {\nabla _{{\bf{\hat x}}}}\log {{q'}_{{\alpha _{t'}}}}\left( {{\bf{\hat x}}\left| {\bf{x}} \right.} \right)} \right\|_2^2} \right].
    \end{align} 
    \end{figure*}
After training, the denoising network can start from ${{\bf{x}}_{T'}} \sim {{\cal N}}\left( {{\bf{0}},{\bf{I}}} \right)$ and generate safeguarding signal samples based on the generation conditions by following the reverse Markov chain, which can be recorded as 
\begin{equation}
\begin{aligned}\label{eq30}
{{\bf{x}}_{t' - 1}} = \frac{1}{{\sqrt {1 - {\beta _{t'}}} }}\left[ {{{\bf{x}}_{t'}} + {\beta _{t'}}{{\bm{\mu }}_{{\bf{\theta ''}}}}\left( {{{\bf{x}}_{t'}},t',{\bf{D'}}} \right)} \right] + \sqrt {{\beta _{t'}}} {{\bf{v}}_{t'}},
\end{aligned} 
\end{equation}
where $t' = T', T' - 1, \ldots ,1$ and ${{\bf{v}}_{t'}} \sim{{\cal N}}\left( {{\bf{0}},{\bf{I}}} \right)$. The safeguarding signals generated here are similar to, but distinct from, those in the training dataset. This ensures a degree of diversity and randomness in the safeguarding signals, preventing unauthorized Rxs from replicating the process. 

\subsubsection{Safeguarding Signal Design and Dataset Construction}
To ensure the aforementioned continuous conditional diffusion model can generate effective safeguarding signals, it is crucial to construct a dataset to train the diffusion model. For instance, we can build a dataset that includes multiple ${1 \mathord{\left/
 {\vphantom {1 {\left| {{H_{wh}}\left( {{f_n},t} \right)} \right|}}} \right.
 \kern-\nulldelimiterspace} {\left| {{H_{wh}}\left( {{f_n},t} \right)} \right|}}$ for waving hand. Using this dataset, the diffusion model is trained to generate safeguarding signals, thereby protecting the user's gestures from unauthorized surveillance. However, in practice, an ISAC system cannot predict user's actions and generate corresponding safeguarding signals. Therefore, this paper designs a composite safeguarding signal and constructs the corresponding dataset to train the diffusion model, ensuring effective protection for users. 

\begin{figure}[t]
\centering
\includegraphics[height=4.5cm]{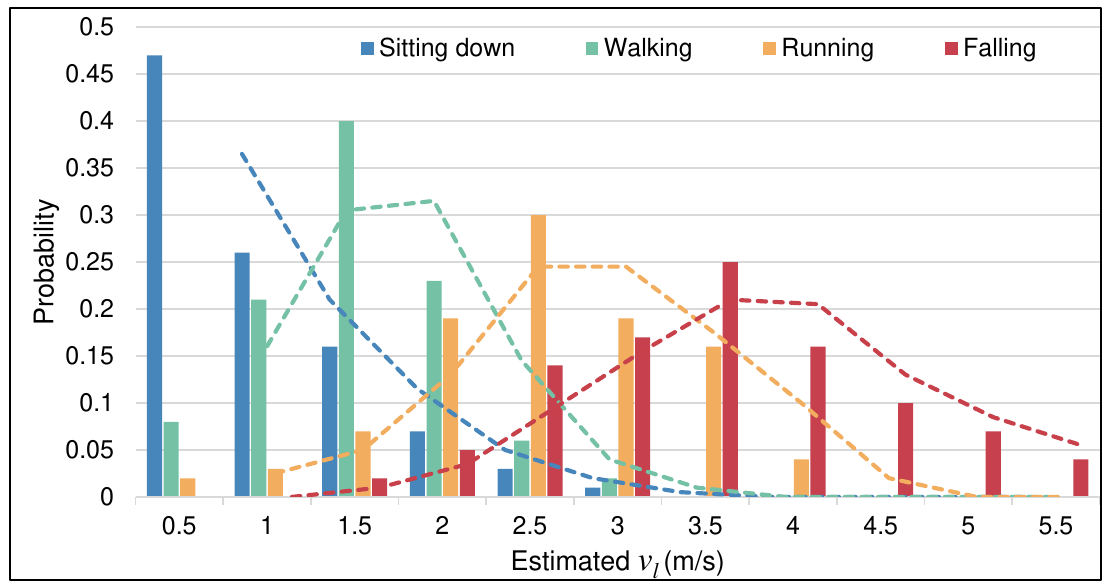} 
\caption{The $v_l$ of different activities in typical indoor scenario.} 
\label{SPD} 
\end{figure}

Specifically, existing research demonstrates that different user activities cause the distinct rate of length
change ${v_l}$. We conducted experiments in a typical indoor scenario based on the method described in~\cite{hu2021defall}, and performed statistical analysis on ${v_l}$ corresponding to four typical user activities. The results are shown in Fig.~\ref{SPD}. As can be seen, for the sitting down, the range of ${v_l}$ extends from approximately 0.2 m/s to 1 m/s, while for falling, it ranges from about 0.8 m/s to 1.5 m/s, which is greater than the other three activities. Given the range of ${v_l}$ caused by indoor activities is limited (approximately 0 to 5.5 m/s) and the purpose of the safeguarding signal is to mask the signal fluctuations caused by user activities, this paper constructs a dataset with multiple composite safeguarding signals for training the continuous diffusion model. Here, each composite safeguarding signal is denoted as 
\begin{equation}
\begin{aligned}\label{eq31}
s\left( {{f_n},t} \right) & = {1 \mathord{\left/
 {\vphantom {1 {\left| {{H_{st}}\left( {{f_n},t} \right)} \right|}}} \right.
 \kern-\nulldelimiterspace} {\left| {{H_{st}}\left( {{f_n},t} \right)} \right|}} + {1 \mathord{\left/
 {\vphantom {1 {\left| {{H_{wk}}\left( {{f_n},t} \right)} \right|}}} \right.
 \kern-\nulldelimiterspace} {\left| {{H_{wk}}\left( {{f_n},t} \right)} \right|}} \\
 &+ {1 \mathord{\left/
 {\vphantom {1 {\left| {{H_{rn}}\left( {{f_n},t} \right)} \right|}}} \right.
 \kern-\nulldelimiterspace} {\left| {{H_{rn}}\left( {{f_n},t} \right)} \right|}} + {1 \mathord{\left/
 {\vphantom {1 {\left| {{H_{fl}}\left( {{f_n},t} \right)} \right|}}} \right.
 \kern-\nulldelimiterspace} {\left| {{H_{fl}}\left( {{f_n},t} \right)} \right|}},
\end{aligned} 
\end{equation}
where ${1 \mathord{\left/
 {\vphantom {1 {\left| {{H_{st}}\left( {{f_n},t} \right)} \right|}}} \right.
 \kern-\nulldelimiterspace} {\left| {{H_{st}}\left( {{f_n},t} \right)} \right|}}$, ${1 \mathord{\left/
 {\vphantom {1 {\left| {{H_{wk}}\left( {{f_n},t} \right)} \right|}}} \right.
 \kern-\nulldelimiterspace} {\left| {{H_{wk}}\left( {{f_n},t} \right)} \right|}}$, ${1 \mathord{\left/
 {\vphantom {1 {\left| {{H_{rn}}\left( {{f_n},t} \right)} \right|}}} \right.
 \kern-\nulldelimiterspace} {\left| {{H_{rn}}\left( {{f_n},t} \right)} \right|}}$, and ${1 \mathord{\left/
 {\vphantom {1 {\left| {{H_{fl}}\left( {{f_n},t} \right)} \right|}}} \right.
 \kern-\nulldelimiterspace} {\left| {{H_{fl}}\left( {{f_n},t} \right)} \right|}}$ are the safeguarding signal corresponding to siting down, walking, running and falling, respectively.

\begin{figure}[t]
\centering
\includegraphics[height=7.5cm]{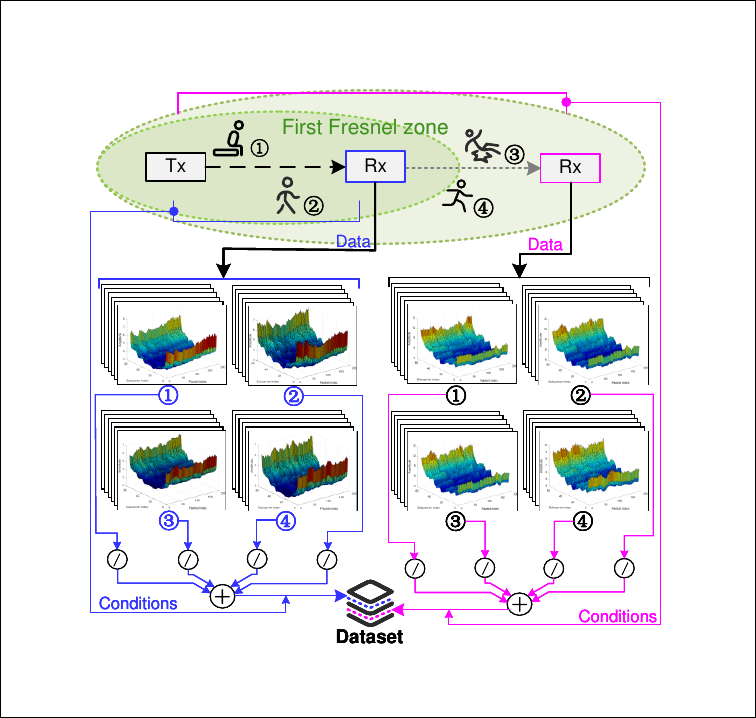} 
\caption{The dataset construction process. Based on the CSI corresponding to four types of activities collected under various conditions, we calculate $s$ via~(\ref{eq31}). Using conditions as labels, we aggregate multiple computed $s$ values to construct the training dataset.} 
\label{DSC} 
\end{figure}

 The reason for selecting these activities to construct the composite safeguarding signal is that ${v_l}$ corresponding to these activities ranges from approximately 0 m to 5.5 m~\cite{wang2017device}, covering the almost all possible ranges in indoor scenarios. Therefore, using $s\left( {{f_n},t} \right)$ yields two distinct advantages: a) $s\left( {{f_n},t} \right)$ can effectively cancel signal fluctuations caused by most user activities in indoor scenarios; b) $s\left( {{f_n},t} \right)$ can introduce certain interference, further masking the signal characteristics. For example, in the case of waving hand, according to the possible range of ${v_l}$, the components ${1 \mathord{\left/
 {\vphantom {1 {\left| {{H_{st}}\left( {{f_n},t} \right)} \right|}}} \right.
 \kern-\nulldelimiterspace} {\left| {{H_{st}}\left( {{f_n},t} \right)} \right|}}$ and ${1 \mathord{\left/
 {\vphantom {1 {\left| {{H_{wk}}\left( {{f_n},t} \right)} \right|}}} \right.
 \kern-\nulldelimiterspace} {\left| {{H_{wk}}\left( {{f_n},t} \right)} \right|}}$ in $s\left( {{f_n},t} \right)$ can partially cancel the signal fluctuations. At the same time, ${1 \mathord{\left/
 {\vphantom {1 {\left| {{H_{rn}}\left( {{f_n},t} \right)} \right|}}} \right.
 \kern-\nulldelimiterspace} {\left| {{H_{rn}}\left( {{f_n},t} \right)} \right|}}$ and ${1 \mathord{\left/
 {\vphantom {1 {\left| {{H_{fl}}\left( {{f_n},t} \right)} \right|}}} \right.
 \kern-\nulldelimiterspace} {\left| {{H_{fl}}\left( {{f_n},t} \right)} \right|}}$ can introduce additional interference, thereby further obscuring the signal fluctuation characteristics. To construct the dataset for training, we utilize the Universal Software Radio Peripheral (USRP) N321 to collect the CSI corresponding to the four types of user activities under various conditions. Subsequently, we compute ${1 \mathord{\left/
 {\vphantom {1 {\left| {{H_{st}}\left( {{f_n},t} \right)} \right|}}} \right.
 \kern-\nulldelimiterspace} {\left| {{H_{st}}\left( {{f_n},t} \right)} \right|}}$, ${1 \mathord{\left/
 {\vphantom {1 {\left| {{H_{wk}}\left( {{f_n},t} \right)} \right|}}} \right.
 \kern-\nulldelimiterspace} {\left| {{H_{wk}}\left( {{f_n},t} \right)} \right|}}$, ${1 \mathord{\left/
 {\vphantom {1 {\left| {{H_{rn}}\left( {{f_n},t} \right)} \right|}}} \right.
 \kern-\nulldelimiterspace} {\left| {{H_{rn}}\left( {{f_n},t} \right)} \right|}}$, and ${1 \mathord{\left/
 {\vphantom {1 {\left| {{H_{fl}}\left( {{f_n},t} \right)} \right|}}} \right.
 \kern-\nulldelimiterspace} {\left| {{H_{fl}}\left( {{f_n},t} \right)} \right|}}$, and then sum them to obtain the composite safeguarding signal as expressed in (\ref{eq31}). Figure~\ref{DSC} illustrates the dataset construction process.

 \subsection{Safeguarding Signal Modulation}
\begin{figure*}[t]
\centering
\includegraphics[width=1\textwidth]{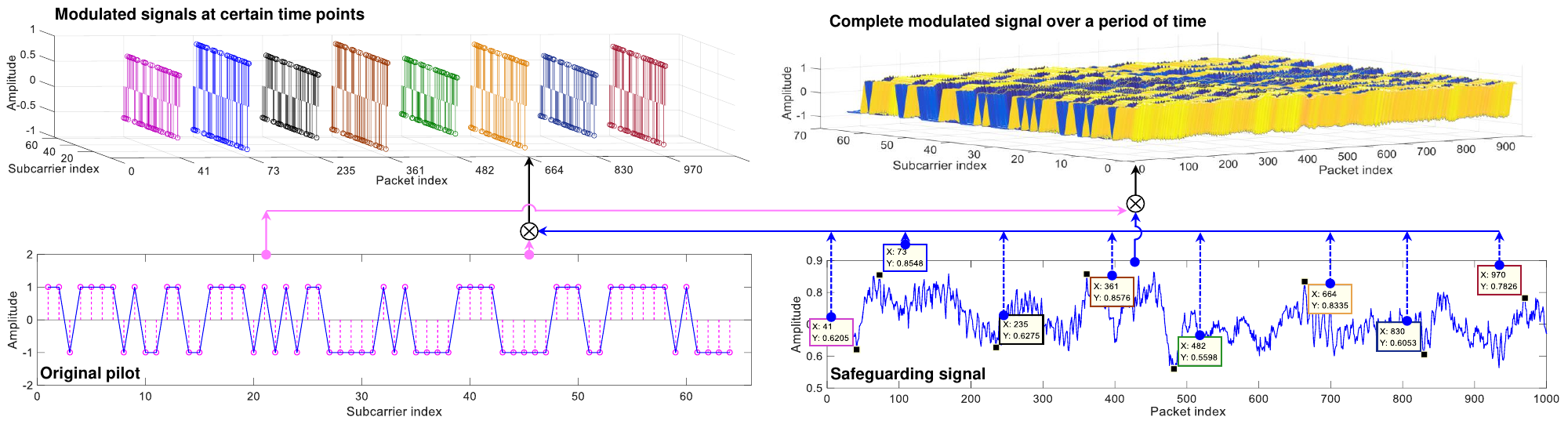}%
\caption{The process of modulating the safeguarding signal onto the original pilot. Here, we show how safeguarding signals are modulated at some certain time points and over a period of time.}
\label{MODU}
\end{figure*}

 Based on the constructed dataset, the continuous diffusion model is trained so that, after training, it can generate composite safeguarding signals under the given conditions during operation. These generated signals possess the same distribution as those in the dataset and exhibit certain diversity and randomness, thereby preventing unauthorized receivers from replicating this process to generate identical signals. Then, the generated signals are modulated onto the real part of the pilot signal. Let the generated protected signal be 
 \begin{align}\label{eq32}
{\bf{s}} = \left[ {s\left( {{f_n},1} \right), \ldots ,s\left( {{f_n},w} \right), \ldots ,s\left( {{f_n},W} \right)} \right],
\end{align}
where $W$ is the length of the safeguarding signal, and the original pilot signal is $\pm 1 + 0{\rm{i}}$, then the pilot signal in the data packet transmitted at time $w$ is $\pm 1 \times s\left( {{f_n},w} \right) + 0{\rm{i}}$, so that 
\begin{align}\label{eq33}
{{\bf{X'}}_{w}} = {\rm{diag}}\left( {s\left( {{f_n},w} \right)X\left[ 0 \right], \ldots ,s\left( {{f_n},w} \right)X\left[ {N - 1} \right]} \right).
\end{align}
This indicates that the safeguarding signals assigned to pilots at different frequencies are the same at time $w$. In this way, the amplitude of the pilot signal in the time domain varies in accordance with the trend represented by the generated safeguarding signal, thereby masking the fluctuation characteristics caused by user activities. Leveraging the same model and random seed, authorized Rx can generate the same safeguarding signal ${\bf{s}}$ as the Tx and multiply it with the original pilot signal to obtain ${\bf{X'}}$. This allows for effective estimation of ${\bf{\hat H}}$, used for communication and sensing, according to (\ref{eq1})-(\ref{eq3}). However, unauthorized receivers can only estimate channels using the original pilot signal, resulting in ${\bf{\hat H'}} = {{\bf{X}}^{ - 1}}{\bf{Y'}}$, which prevents them from illegitimate sensing. Figure~\ref{MODU} shows the process of modulating the generated signal onto the pilot.
\section{Implementation and Evaluation}
In this section, we utilize the USRP to evaluate the proposed method, covering three aspects. First, we analyze the graph generation for the link and nodes activation. Next, we evaluate the safeguarding signal generation and modulation. Finally, the performance in protecting users from unauthorized monitoring is assessed, using human activity recognition as a case study.
\subsection{Experimental design }
\subsubsection{Experimental Configurations}
\begin{figure*}[t]
\centering
\includegraphics[width=1\textwidth]{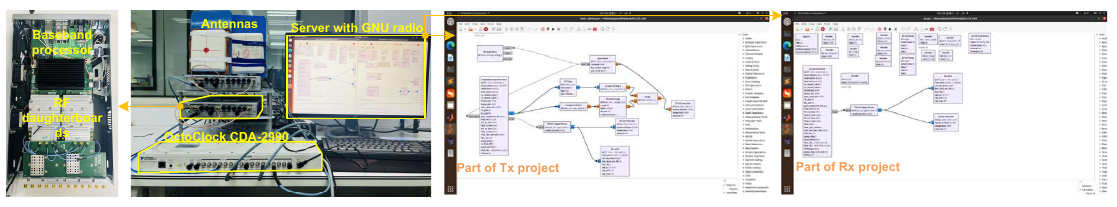}%
\caption{The hardware equipment and GNUradio projects for experiments. Here, we display the server, USRP N321 (and its internal structure), an external clock, and the used directional antenna. Certain parts of the GNUradio project for Tx and Rx are also presented.}
\label{HDW}
\end{figure*}

We conduct experiments using servers and USRP N321 devices. Specifically, the training and inference of the diffusion model are performed on a server running the Ubuntu 20.04 operating system, equipped with the AMD Ryzen Threadripper PRO 3975WX 32-core processor and the NVIDIA RTX A5000 GPU. The generated signals are then transmitted to another server, which has 64 GB of memory and GNU Radio~\footnote{https://www.gnuradio.org/}, and connects to USRP devices via optical fibers and 10-gigabit Ethernet cables. Each N321 device\footnote{https://www.ettus.com/all-products/usrp-n321/} includes the Xilinx Zynq-7100 SoC based baseband processor and UBX-160 daughterboard\footnote{https://www.ettus.com/all-products/ubx160/}, supporting up to 200 MHz bandwidth. The devices are synchronized using the OctoClock-G CDA-2990\footnote{https://www.ettus.com/all-products/octoclock-g/}. During the experiments, the signal center frequency is set at 2.8 GHz, with a bandwidth of 100 MHz covered by 64 subcarriers. We adopt a block-type pilot arrangement and transmit data at the rate of 100 packets per second. Additionally, to ensure the quality of the training data, the directional antenna with a gain of 12 dBi is used for transmitting and receiving signals. Figure~\ref{HDW} displays the prototype diagram and part of the GNU projects.
\subsubsection{Experimental Method} We first train the proposed D-CDM, where the input includes device locations and user location, and the output is the generated graph depicting the wireless links, Tx, and Rx that need to be activated. This generated graph is evaluated by reward function, and the feedback is used to optimize the network parameters to complete the training. The continuous diffusion model is then trained with the dataset that contains composite safeguarding signals. Note that each composite safeguarding signal in the dataset is labeled with the locations of the Tx, Rx, and user, which are used as training conditions. During the testing, we first put the spatial distribution of ISAC devices and user location into the trained discrete diffusion model to analyze the graph generation performance, including the generation process and final outcomes. After that, the activated links, nodes, and user location are fed to continuous diffusion model as conditions to analyze the safeguarding signal generation. Finally, we modulate the generated signals onto the pilot and use the activity recognition methods described in~\cite{zhang2022csi},~\cite{chen2018wifi},~\cite{huang2020towards}, and ~\cite{lu2022cehar} as an examples to evaluate the system performance in protecting users from unauthorized sensing.

\subsubsection{Evaluation Metrics} For the generated safeguarding signals, we use the structural similarity index measure (SSIM) and Fréchet inception distance (FID) to evaluate the similarity between the generated signals and those in the training dataset. SSIM measures similarity by analyzing the mean and covariance between two samples, with higher values indicating greater similarity. FID assesses similarity by measuring the Fréchet distance between the high-level features of the generated and training signals, where lower scores indicate greater similarity. For activity recognition, the recognition accuracy (RA) and the accuracy degradation rate (ADR) are used to evaluate the system performance. Here, RA is calculated by dividing the number of correct recognitions by the total number of tests. The ADR is defined as $ADR = {{\left( {A{c_{org}} - A{c_{sf}}} \right)} \mathord{\left/
 {\vphantom {{\left( {A{c_{org}} - A{c_{sfg}}} \right)} {A{c_{org}}}}} \right.
 \kern-\nulldelimiterspace} {A{c_{org}}}}$, where $A{c_{org}}$ is the RA without safeguarding signal, and $A{c_{sf}}$ is the RA of unauthorized APs after the introduction of safeguarding signals. A higher ADR means that the proposed method offers better protection for the user.
\subsection{Experimental Results}
\subsubsection{Graph Generation}
\begin{figure}[t]
\centering
\includegraphics[height=4.5cm]{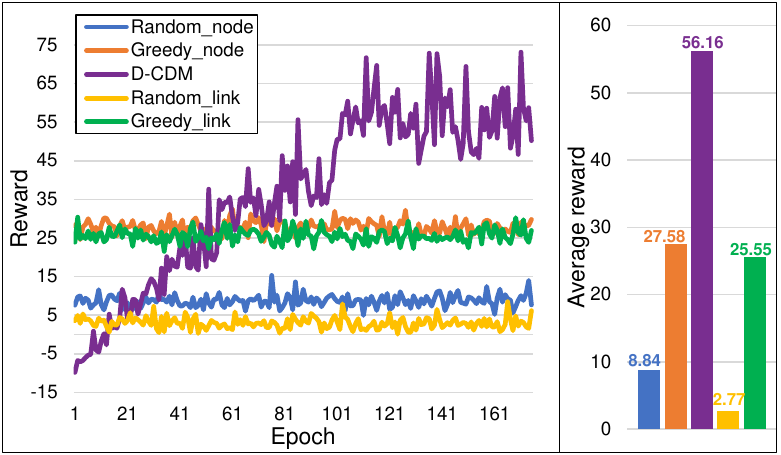} 
\caption{The reward of D-DCM and other baselines over training epochs.} 
\label{TR} 
\end{figure}

First, we analyze the graph generation performance of the proposed D-CDM. Figure~\ref{TR} shows the training curve of D-CDM  and compares it with other four baselines. Here, the node-based and link-based greedy methods activate the four nodes and links closest to the user for sensing, respectively, while the random methods activate nodes and links randomly. As can be seen, D-CDM converges after roughly 100 epochs with an average reward of about 56.16, indicating effective network parameter optimization and generation strategy learning via feedback from the reward function. In contrast, the average rewards for node-based and link-based random methods are approximately 8.84 and 2.77, respectively, and the greedy approaches yield rewards of 27.58 and 25.55, respectively. This can be interpreted by that D-CDM generates graphs from random noise, introducing a degree of randomness. This randomness endows D-CDM with superior exploratory capabilities, allowing it to find solutions with better reward. Moreover, the D-CDM considers the impact of both nodes and edges on the reward during the denoising process, making the denoising more comprehensive and generating graphs that better meet the requirements.

\begin{figure*}[t]
\centering
\includegraphics[width=1\textwidth]{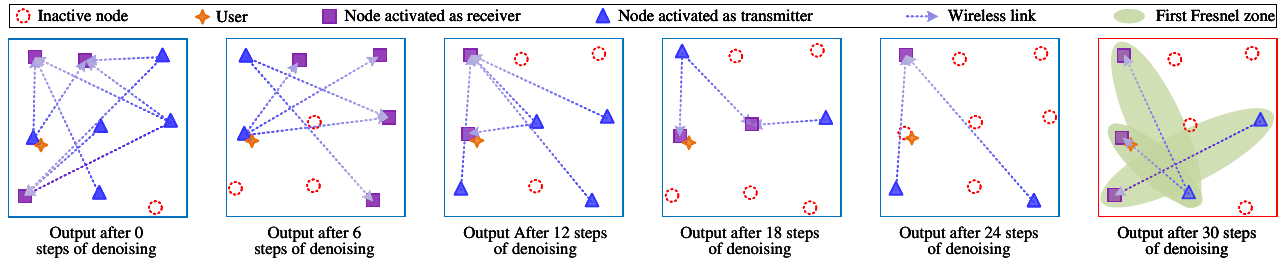}%
\caption{The graph generation process based on trained D-CDM.}
\label{GRF-G}
\end{figure*}

\begin{figure*}[t]
\centering
\includegraphics[width=1\textwidth]{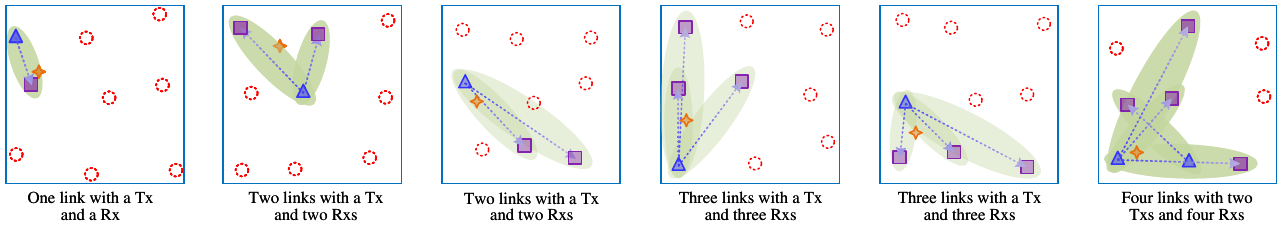}%
\caption{The graph generation results under different generative conditions, i.e., different ISAC device distributions.}
\label{GRF-R}
\end{figure*}

Second, Fig.~\ref{GRF-G} depicts the graph generation process by the trained D-CDM under various input conditions. As presented, when the number of steps is less than 12, the denoising is incomplete, resulting in graphs populated with noisy edges and nodes, which may be useless for effective sensing. As the denoising steps increase, D-CDM continuously refines the noisy graph by appropriately adding or removing nodes and edges. After 18 steps, there is a noticeable decrease in the number of ineffective edges and nodes. Ultimately, an optimal graph is produced after 30 denoising steps, as indicated by the results in the red box. In this graph, the activated links form a Fresnel zone around the user, allowing the ISAC system to sense more efficiently. Figure~\ref{GRF-R} displays the generated results under different spatial distributions of ISAC devices. The generated graphs indicate that the trained D-CDM can adapt the node and edge generations according to the inputs, hence generating the desired graphs under various conditions, demonstrating the generalizability of the proposed D-CDM. Additionally, the optimal graph generated with 4 wireless links includes 2 Txs and 4 Rxs, which is more energy-efficient than having 3 Txs and 4 Rxs, as using more Txs consumes more energy. This further validates the rationality of the D-CDM. 

\subsubsection{Safeguarding Signal Generation}
\begin{figure*}[htbp]
\centering
\subfigure[The result after 400 steps of denoising.]{
\begin{minipage}[t]{0.3\linewidth}
\centering
\includegraphics[width=4.5cm]{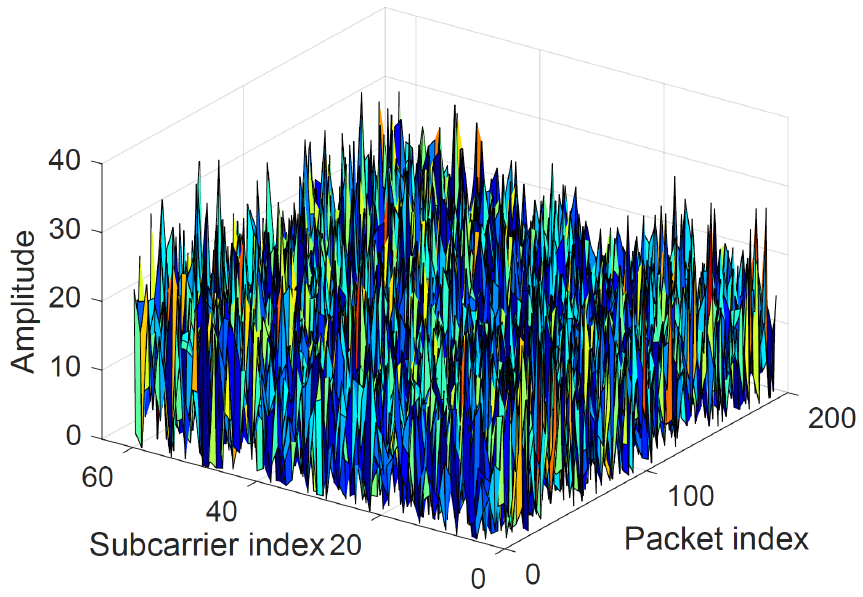}
\end{minipage}%
}%
\subfigure[The result after 460 steps of denoising.]{
\begin{minipage}[t]{0.35\linewidth}
\centering
\includegraphics[width=4.5cm]{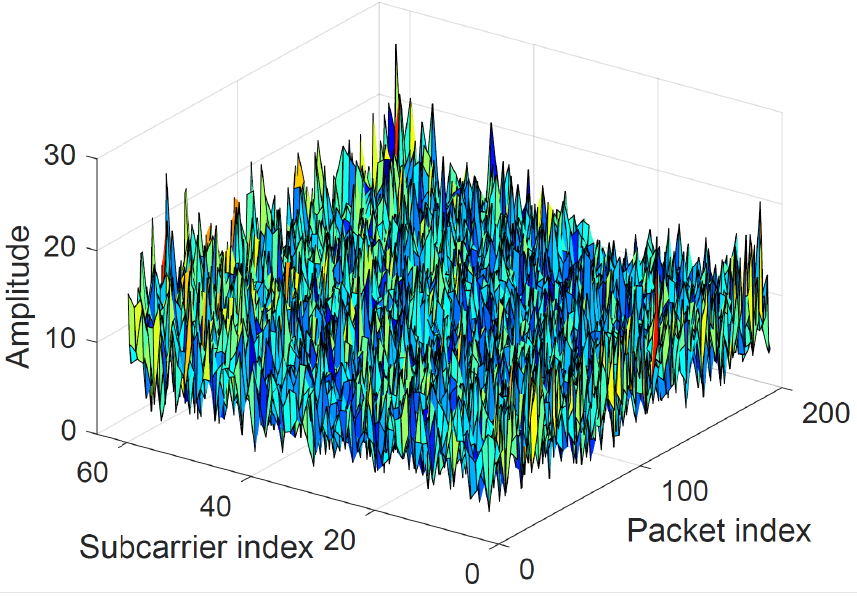}
\end{minipage}%
}%
\subfigure[The result after 480 steps of denoising.]{
\begin{minipage}[t]{0.3\linewidth}
\centering
\includegraphics[width=4.5cm]{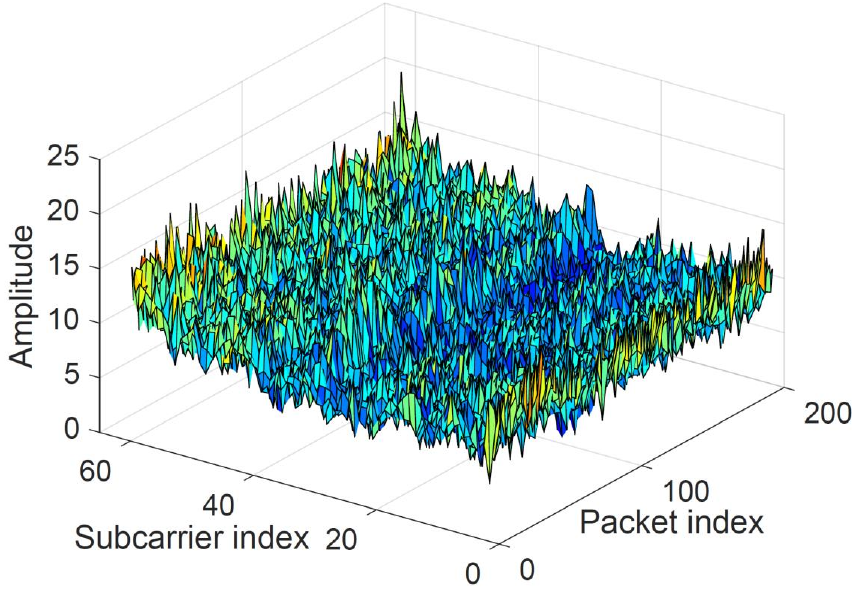}
\end{minipage}
} \\
\subfigure[The result after 490 steps of denoising.]{
\begin{minipage}[t]{0.3\linewidth}
\centering
\includegraphics[width=4.5cm]{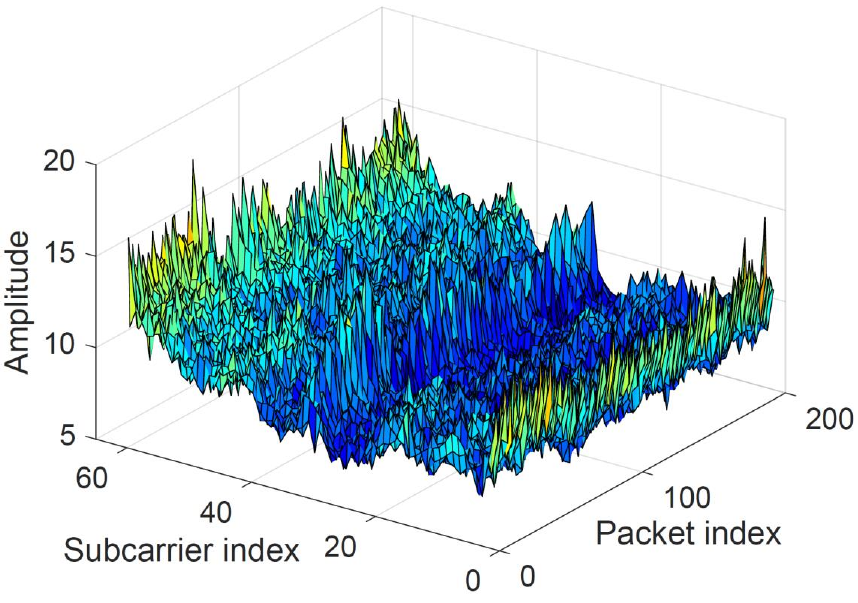}
\end{minipage}%
}%
\subfigure[The result after 500 steps of denoising.]{
\begin{minipage}[t]{0.35\linewidth}
\centering
\includegraphics[width=4.5cm]{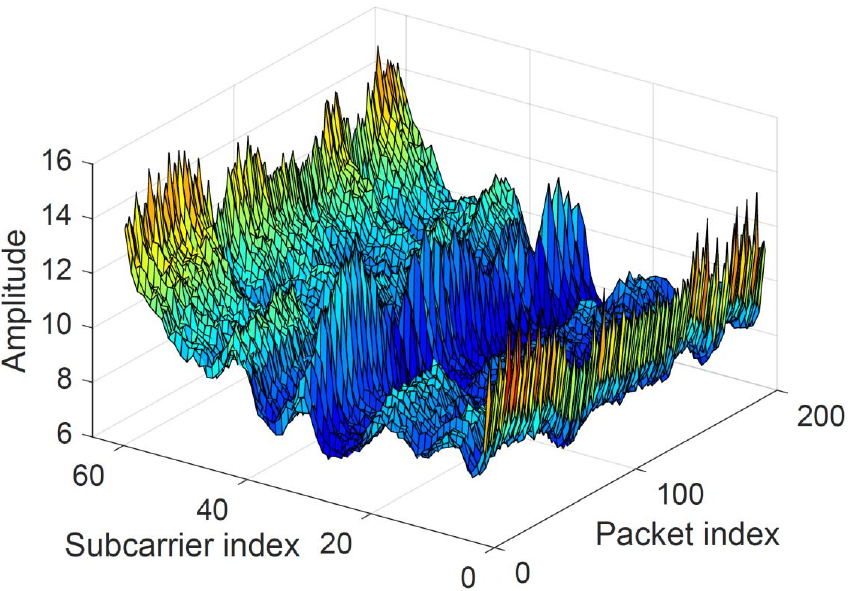}
\end{minipage}%
}%
\subfigure[The training data from the dataset.]{
\begin{minipage}[t]{0.3\linewidth}
\centering
\includegraphics[width=4.5cm]{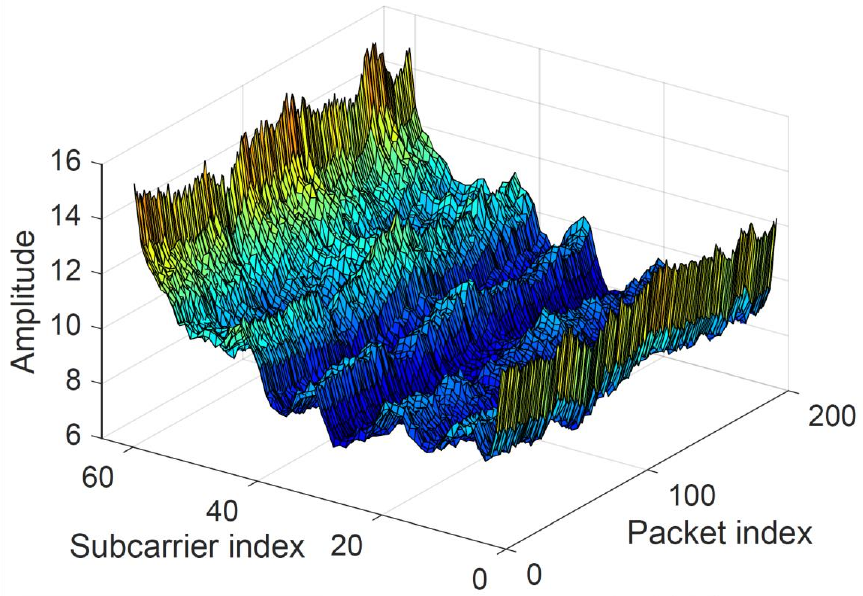}
\end{minipage}
}%
\centering
\caption{ The safeguarding signal generation process when Tx and Rx are about 6 meters apart. }
\label{F8}
\end{figure*}

\begin{figure*}[htbp]
\centering
\subfigure[The result after 400 steps of denoising.]{
\begin{minipage}[t]{0.3\linewidth}
\centering
\includegraphics[width=4.5cm]{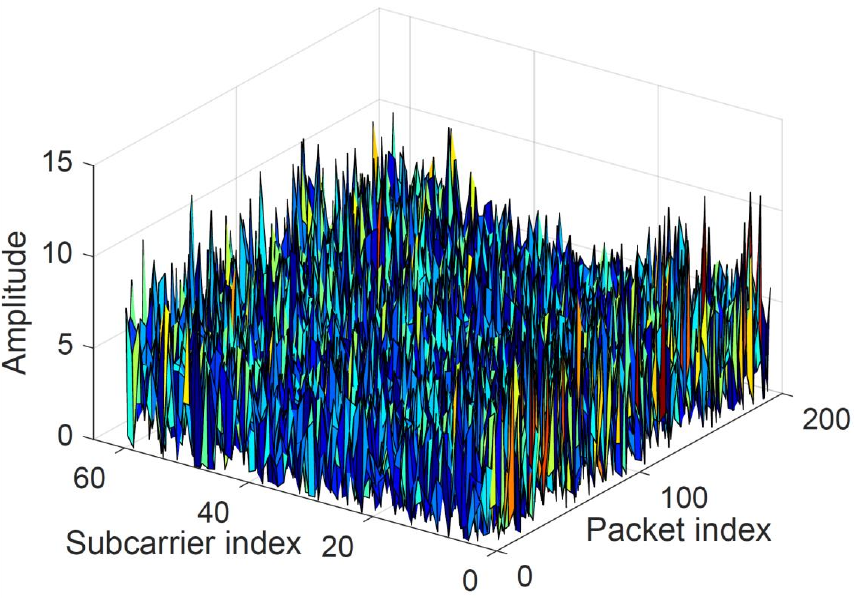}
\end{minipage}%
}%
\subfigure[The result after 460 steps of denoising.]{
\begin{minipage}[t]{0.35\linewidth}
\centering
\includegraphics[width=4.5cm]{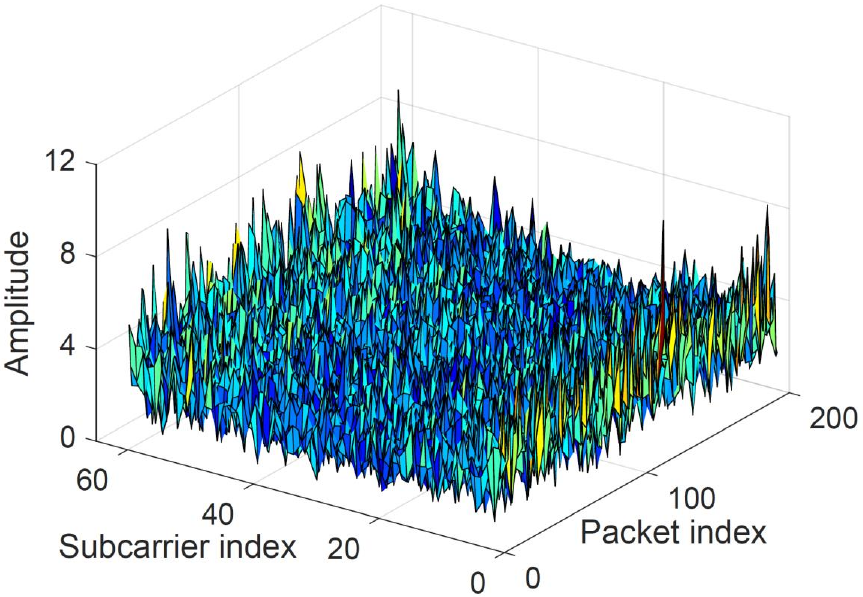}
\end{minipage}%
}%
\subfigure[The result after 480 steps of denoising.]{
\begin{minipage}[t]{0.3\linewidth}
\centering
\includegraphics[width=4.5cm]{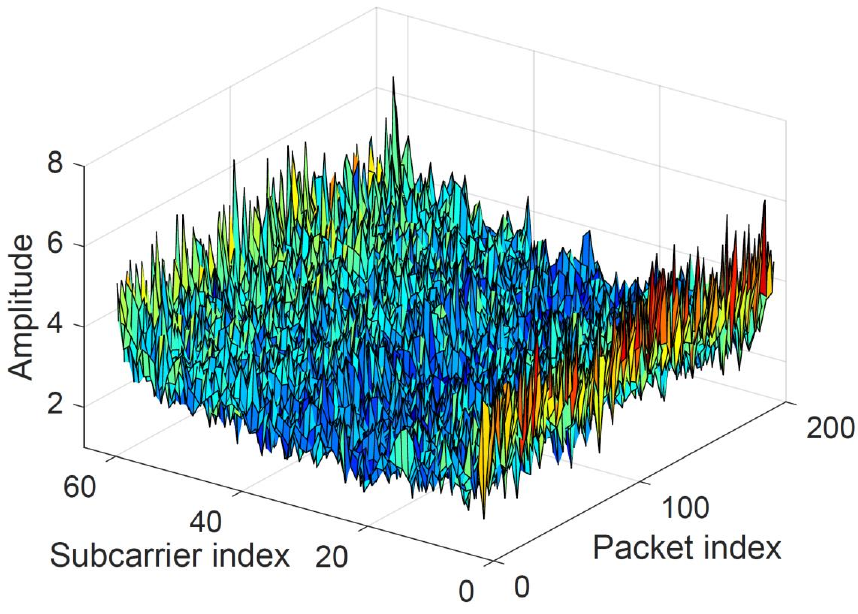}
\end{minipage}
} \\
\subfigure[The result after 490 steps of denoising.]{
\begin{minipage}[t]{0.3\linewidth}
\centering
\includegraphics[width=4.5cm]{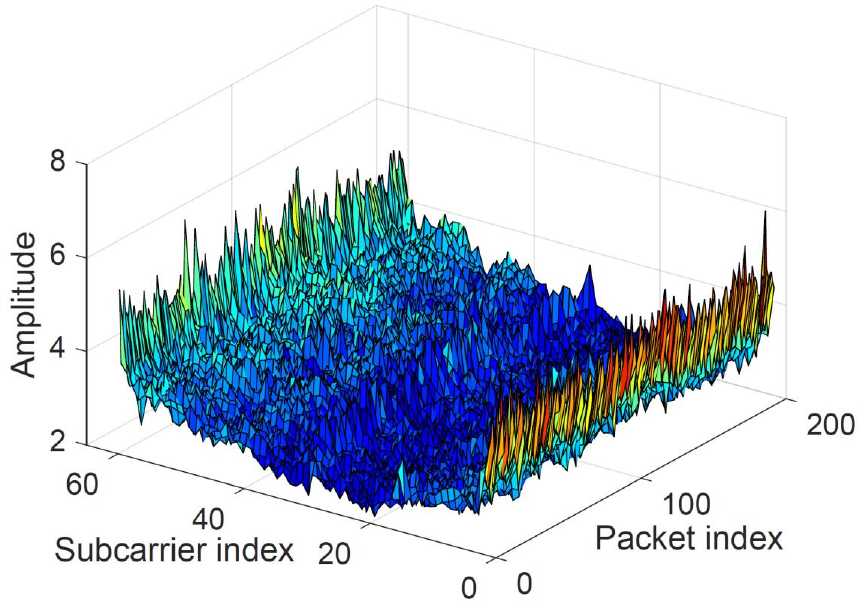}
\end{minipage}%
}%
\subfigure[The result after 500 steps of denoising.]{
\begin{minipage}[t]{0.35\linewidth}
\centering
\includegraphics[width=4.5cm]{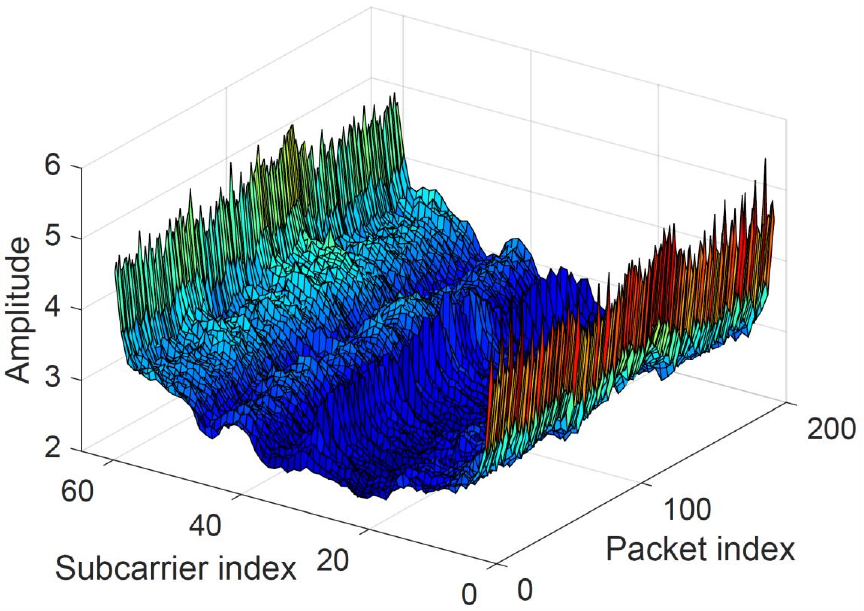}
\end{minipage}%
}%
\subfigure[The training data from the dataset.]{
\begin{minipage}[t]{0.3\linewidth}
\centering
\includegraphics[width=4.5cm]{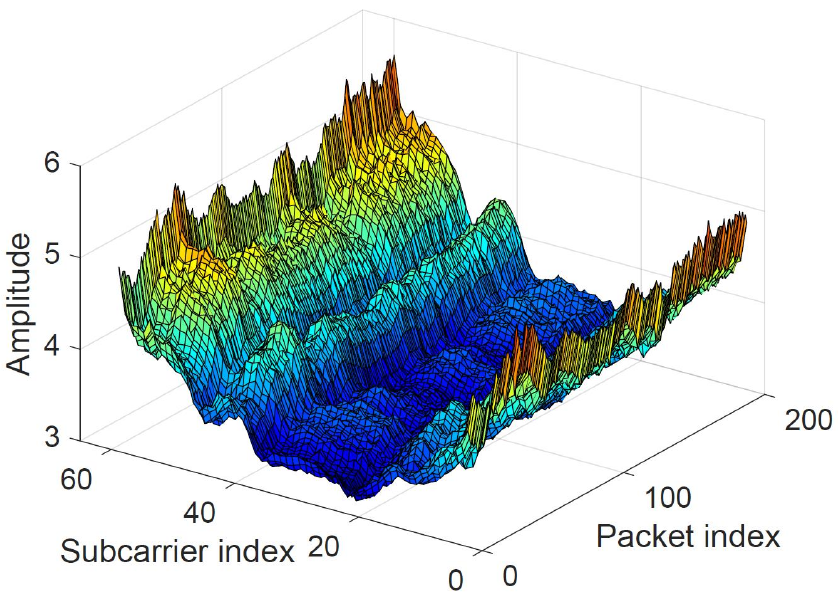}
\end{minipage}
}%
\centering
\caption{The safeguarding signal generation process when Tx and Rx are about 3 meters apart. }
\label{F9}
\end{figure*}

Using the activated links, nodes, and user location as generation conditions, we further analyze safeguarding signal generation. Figures~\ref{F8} and~\ref{F9} display the generation process under various conditions. As shown, starting with random noise, when the number of steps is less than 460, the generated signals are noisy, indicating insufficient denoising. As denoising continues, noise is further removed, making the safeguarding signals clearer and more structured. After completing 500 steps, we obtain the generated safeguarding signals, which exhibit similar trends to the training data but are distinct from that, confirming the effectiveness of the safeguarding signal generation.

Additionally, different input conditions yield unique safeguarding signals, particularly in signal strength and trend. For instance, when the Tx and Rx in the generation conditions are closer, the resulting safeguarding signals are weaker, whereas they are stronger when Tx and Rx are further apart. This happens because the Rx captures signals with higher amplitude and clear fluctuation characteristics when Tx and Rx are near each other, necessitating weaker safeguarding signals to effectively mask fluctuations caused by user activities. Conversely, a greater distance between Tx and Rx demands a relatively stronger safeguarding signal. Building on this, we compare the generated safeguarding signals with the training data. The results show that while the generated signals closely match the training data in intensity and overall trends, they remain distinct. This highlights the diversity and randomness of the signals produced by the diffusion model, complicating replication by unauthorized devices, thereby providing strong protection for the user.


\begin{figure}[tb]
\centering
\includegraphics[height=3.2cm]{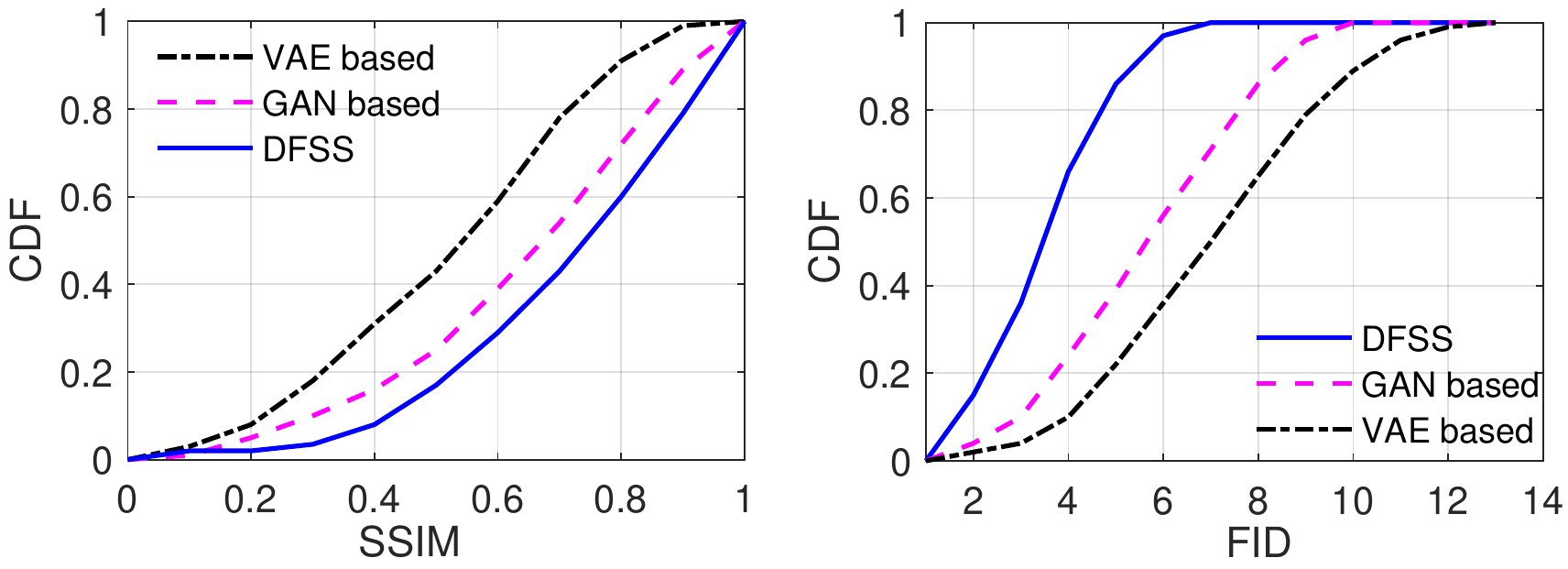} 
\caption{The quality of the generated safeguarding signal.} 
\label{F10} 
\end{figure}

Subsequently, we evaluate the quality of the generated signals using the FID and SSIM metrics, and compare the proposed DFSS with other two methods. The results, presented in Fig.~\ref{F10}, show that DFSS achieves a median SSIM of 0.73. In comparison, methods based on GAN ~\cite{radford2015unsupervised} and VAE ~\cite{sohn2015learning} achieve median SSIMs of 0.68 and 0.57, respectively. Additionally, the median FID of DFSS is 3.5, which is lower than 5.8 and 7.1 of GAN-based and VAE-based methods, respectively. Such performance of DFSS can be attributed to two main factors. First, the diffusion model in DFSS features strong exploratory capabilities and a progressive denoising approach. This allows for thorough exploration in signal generation and helps retain more detailed features of the signal during the generation process. Second, DFSS takes into account the generative conditions during both training and inference. This enables it to produce safeguarding signals that are better tailored to the specified conditions.

\subsubsection{Protection Performance in Activity Recognition}

\begin{figure*}[htbp]
\centering
\subfigure[AF-ACT.]{
\begin{minipage}[t]{0.2425\linewidth}
\centering
\includegraphics[width=4.4cm]{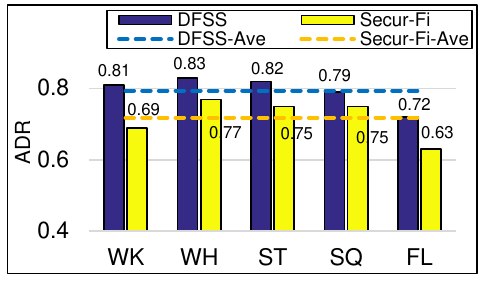}
\end{minipage}%
}%
\subfigure[ABLSTM.]{
\begin{minipage}[t]{0.2415\linewidth}
\centering
\includegraphics[width=4.3cm]{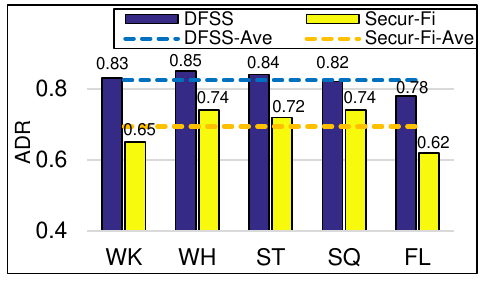}
\end{minipage}%
}%
\subfigure[PhaseAnti.]{
\begin{minipage}[t]{0.2345\linewidth}
\centering
\includegraphics[width=4.3cm]{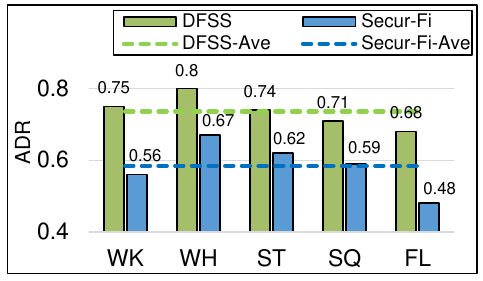}
\end{minipage}
}%
\subfigure[CeHAR.]{
\begin{minipage}[t]{0.245\linewidth}
\centering
\includegraphics[width=4.3cm]{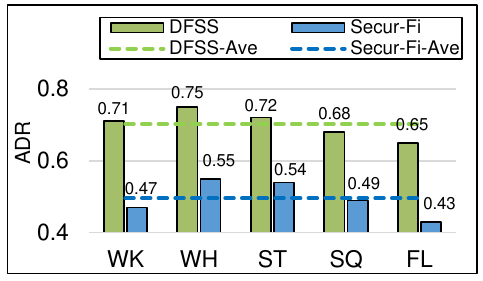}
\end{minipage}
}%
\centering
\caption{The impact of DFSS on user activity recognition for unauthorized devices.}
\label{ADR}
\end{figure*}

Based on the activity recognition methods introduced in~\cite{zhang2022csi},~\cite{chen2018wifi},~\cite{huang2020towards}, and ~\cite{lu2022cehar} (denoted as AF-ACT, ABLSTM, PhaseAnti, and CeHAR, respectively), we analyze the impact of DFSS on recognition accuracy using 5 actions (including walking (WK), waving hand (WH), sitting (ST), squatting (SQ), and falling (FL)) and compare it with Secur-Fi~\cite{meng2023secur}. As shown in Fig.~\ref{ADR}, with DFSS, the average ADR of the these systems can reach 0.82, 0.79, 0.74, and 0.7, respectively. These are higher compared to those achieved with Secur-Fi, which are 0.69, 0.72, 0.58, and 0.50, respectively. This is because DFSS modulates the generated safeguarding signal onto the pilot, directly influencing signal fluctuations. Compared with the Secur-Fi, which introduces interference by adjusting the antenna, our approach affects the signal more thoroughly and directly, thereby achieving better performance.

Additionally, from the perspective of user activities, the influence of DFSS is consistent. For instance, in AF-ACT, the ADR for the five activities are 0.81, 0.83, 0.82, 0.79, and 0.72, respectively, demonstrating DFSS's general applicability across different activities. From the system perspective, the impact of DFSS varies. Specifically, DFSS has a more significant impact on AF-ACT and ABLSTM compared with PhaseAnti and CeHAR. This is because the former two rely solely on CSI amplitude for recognition, while the latter two use both amplitude and phase information, enhancing their robustness. Overall, DFSS effectively declines the performance of unauthorized systems, thereby achieving the goal of user protection.

\begin{figure*}[htbp]
\centering
\subfigure[The confusion matrix of AF-ACT.]{
\begin{minipage}[t]{0.3\linewidth}
\centering
\includegraphics[width=5.5cm]{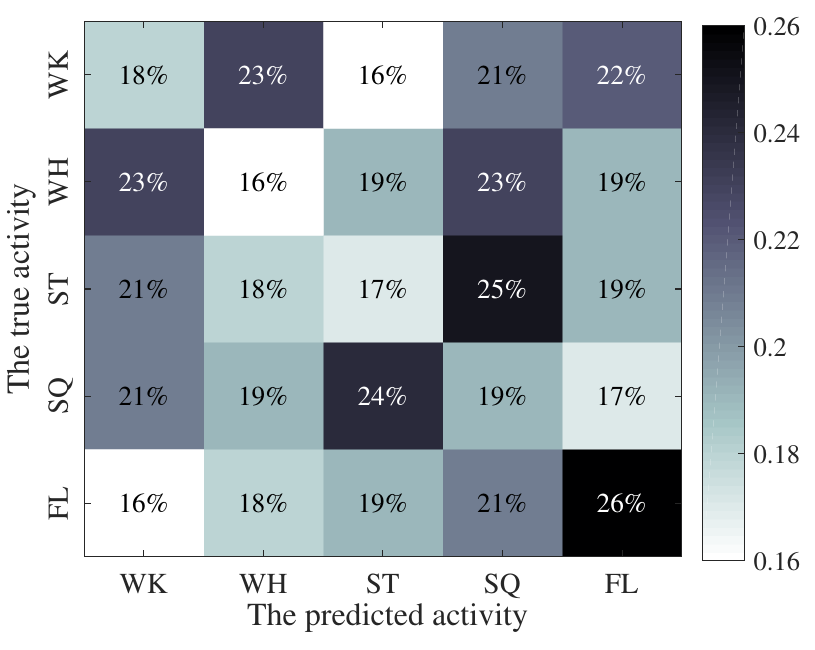}
\end{minipage}%
}%
\subfigure[The confusion matrix of CeHAR.]{
\begin{minipage}[t]{0.35\linewidth}
\centering
\includegraphics[width=5.5cm]{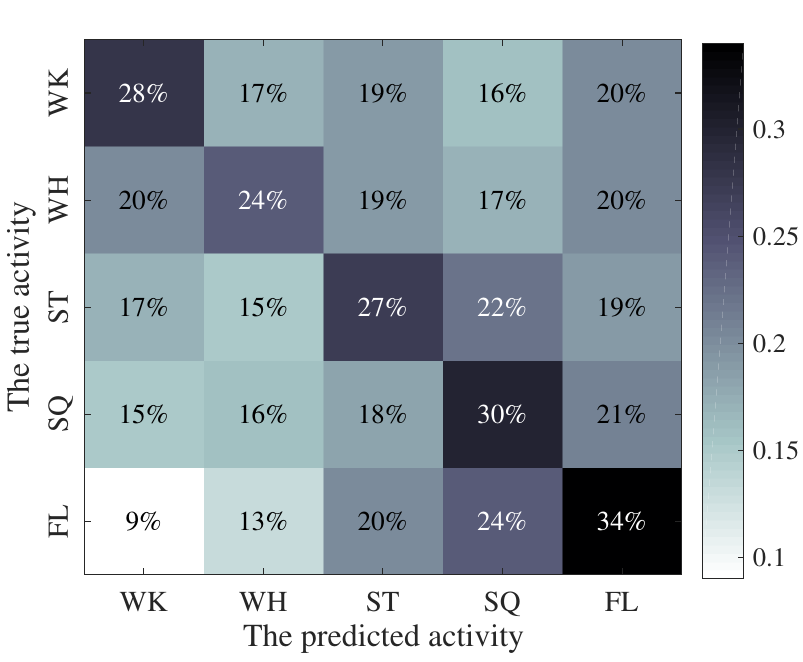}
\end{minipage}%
}%
\subfigure[The impact of communication speed on DFSS.]{
\begin{minipage}[t]{0.3\linewidth}
\centering
\includegraphics[width=5.5cm]{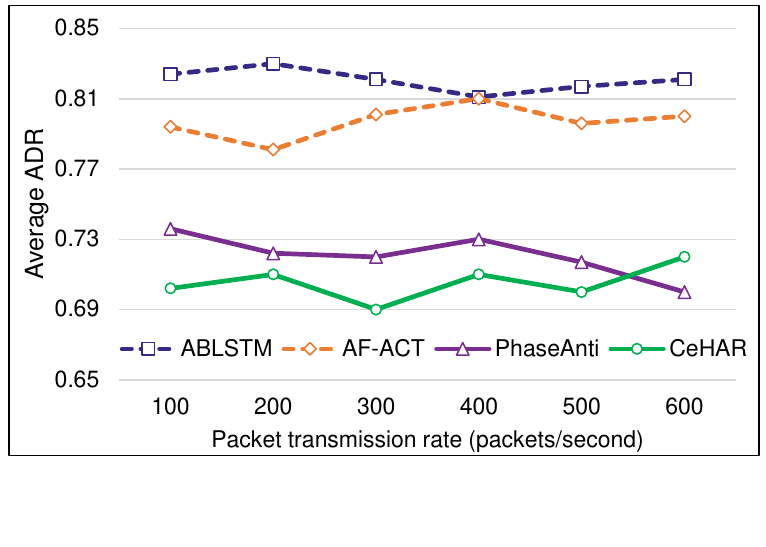}
\end{minipage}
} 
\centering
\caption{The recognition accuracy analysis of AF-ACT and CeHAR, and the impact of communication speed on DFSS performance.}
\label{COF}
\end{figure*}

Using ABLSTM and CeHAR as examples, Figs.~\ref{COF}(a) and (b) show the activity recognition confusion matrices with DFSS applied. Specifically, for five activities, ABLSTM's recognition accuracies are 18\%, 16\%, 17\%, 19\%, and 26\%, while those of CeHAR are 28\%, 24\%, 27\%, 30\%, and 34\%, respectively. This further demonstrates the robust protective capability of DFSS. Importantly, the results indicate that DFSS does not lead the systems to incorrectly classify different activities into the same one. This happens because the generated safeguarding signals are diverse and random, causing the CSI features captured by unauthorized devices to vary, which ultimately results in unpredictable recognition outcomes.

Finally, we explored the impact of communication speed (i.e., data packet transmission rate) on DFSS. The results in Fig.~\ref{COF}(c) reveal that across different transmission rates, the average ADR of each system remains nearly consistent. For example, with data packet transmission rate ranging from 100 to 600, ADRs for AF-ACT are 0.79, 0.78, 0.80, 0.81 0.80, and 0.8, and for CeHAR, they are 0.70, 0.71, 0.69, 0.71, 0.70, and 0.72. This is because the generated safeguarding signals can be resampled to adapt to different communication speeds, ensuring DFSS's performance under varying conditions, which is critical for practical applications.

\section{Conclusion}
We have proposed DFSS, a secure sensing system that generates safeguarding signals and modulates them onto the pilot to mask signal fluctuations caused by user activity, thereby protecting users from unauthorized monitoring. In DFSS, we have developed a discrete conditional diffusion model to generate graphs that guide the ISAC system in activating appropriate links and nodes, ensuring economical operation while maintaining sensing performance. Additionally, we have designed a composite safeguarding signals for typical indoor activities and constructed a dataset to train the continuous conditional diffusion model, enabling it to generate safeguarding signals based on the input conditions. These signals, characterized by diversity and randomness, can effectively mask signal fluctuations features due to user activities, thereby preventing unauthorized devices from extracting actual CSI for illegitimate monitoring. Using activity recognition as an example, the evaluation shows that DFSS can reduce the recognition accuracy of unauthorized devices by approximately 70\%, demonstrating the potential of GAI in enhancing sensing security. Future work will explore the use of GAI for secure user localization and tracking.
\bibliographystyle{IEEEtran}
\bibliography{Ref.bib}

\end{document}